\documentclass[12pt]{article}
\pdfoutput =1
\usepackage{float} 
\usepackage{amsmath}
\usepackage{amsfonts}   
\usepackage{amssymb}

\begin{document}
\date{}
\title{{\bf{\Large Chern-Simons vortices and holography}}}
\author{
 {\bf {\normalsize Dibakar Roychowdhury}$
$\thanks{E-mail:  dibakarphys@gmail.com, dibakar@cts.iisc.ernet.in}}\\
 {\normalsize Centre for High Energy Physics, Indian Institute of Science, }
\\{\normalsize C.V. Raman Avenue, Bangalore 560012, Karnataka, India}
}

\maketitle
\begin{abstract}
In this paper, based on the $ AdS_{4}/CFT_{3} $ duality, we have explored the precise connection between the abelian Chern-Simons (CS) Higgs model in ($ 2+1 $) dimensions to that with its dual gravitational counterpart living in one higher dimension. It has been observed that the $ U(1) $ current computed at the boundary of the $ AdS_{4} $ could be expressed as the local function of the vortex solution that has the remarkable structural similarity to that with the Ginzburg-Landau (GL) type local expression for the current associated with the Maxwell-CS type vortices in ($ 2+1 $) dimensions. In order to explore this duality a bit further we have also computed the coherence length as well as the magnetic penetration depth associated with these vortices. Finally using the knowledge of both the coherence length as well as the magnetic penetration depth we have computed the Ginzburg-Landau coefficient for the Maxwell-CS type vortices in ($ 2+1 $) dimensions.  
 \end{abstract}

\section{Overview and Motivation}

During the past several decades, the Chern-Simons (CS) theories coupled to matter fields had drawn renewed attention due to its various remarkable features among which the existence of charged vortices, popularly known as the CS vortices could be regarded as one of the significant achievements of such theories. In \cite{ref1} it was shown for the first time that the abelian CS Higgs model in ($ 2+1 $) dimensions could give rise to charged vortices of finite energy. These vortices are fundamentally different from that of the usual vortex structure known in the context of Ginzburg-Landau (GL) theory or the abelian Higgs model in the sense that they carry both the electric charge as well as the magnetic flux, which is the reason why these charged objects are considered to be the excellent candidates for the so called \textit{anyons}.

Since the discovery, the various physical properties of these CS vortices have been explored in different directions \cite{ref2}-\cite{ref10}. For example, the self dual CS vortices satisfying a set of self duality equations were first investigated in \cite{ref2}. Apart from these earlier analyses, recently some more attempts have been made from various other perspectives \cite{ref11}-\cite{ref13}, for example, in \cite{ref13} the authors have analytically studied the so called BPS vortex equations for a specific choice of the non canonical kinetic term and found that these BPS vortices exhibit certain properties that are quantitatively different from that of the standard CS vortices.

Beside these advancements, for the past couple of decades, there has been another significant development in the area of theoretical physics which is popularly known as the AdS/CFT correspondence \cite{ref14}-\cite{ref15}. It is now widely believed that the AdS/CFT duality essentially captures all the underlying physics of strongly coupled CFTs. One of the remarkable achievements of AdS/CFT duality is that it provides an exact frame work to study these strongly coupled CFTs at finite temperature. The dual description of such CFTs typically possess charged AdS Reissner-Nordstrom (RN) black brane solutions which eventually become unstable to develop a charged hair below certain critical value of the temperature ($ T_c $).   

These AdS black brane solutions with charged hair essentially describe a new phase in the dual CFTs. As the gauge fields do not have any dynamics at the boundary of the AdS space, therefore these dual CFTs do not have any dynamical gauge bosons either and as a consequence of this the new phase in these dual CFTs should be treated as the super-fluid phase rather than the superconducting phase. In other words, the instability occurring in the AdS RN black brane solution eventually breaks the global $ U (1) $ symmetry for the boundary theory \cite{ref16}-\cite{ref23}. 

During the past couple of years the existence of the triangular lattice configurations for holographic type II superconductors have been explored under various circumstances \cite{Maeda:2009vf}-\cite{Banerjee:2013maa}. It is known from the standard Ginzburg-Landau (GL) theory for type II superconductors that the so called GL current associated with the triangular lattice configuration could be expressed as a local function of the vortex solution circulating around the core of the vortex \cite{ref24}-\cite{ref25}. One of the remarkable outcomes of these holographic calculations is that under the long wavelength approximation, the abelian Higgs model coupled to gravity \cite{ref16} can indeed produce the so called \textit{local} structure for the supercurrent surrounding the vortices \cite{Maeda:2009vf}. This therefore provides an exact holographic framework to study the standard GL theory for type II superconductors.

Following the same spirit as mentioned above, the purpose of the present article is to establish the precise AdS/CFT framework to study the abelian CS Higgs model in ($ 2+1 $) dimensions. In other words, in the present work we want to address the following question: Is it possible to construct certain dual gravitational theory for the abelian CS (Higgs) vortices in ($ 2+1 $) dimensions that can eventually give rise to some \textit{local} GL type current associated with these vortices. To answer this question one should take into account the following issues. As a first step, one should independently calculate the current associated with these CS vortices in ($ 2+1 $) dimensions and see whether it is indeed possible to find a GL (like) \textit{local} expression for the current. The next step would be to search for a dual gravitational theory that is coupled to certain matter fields in $ AdS_{4} $ so that the $ U(1) $ current computed at the boundary of the $ AdS_{4} $ could be eventually brought in to a similar local form. Based on the structural similarities between these two local expressions of the current, one should be able to find the holographic framework that eventually describes the abelian CS Higgs model in ($ 2+1 $) dimensions. 

The organization of the paper and the summary of results are the following: In Section 2, we have considered the standard abelian CS Higgs model in ($ 2+1 $) dimensions and computed the usual Ginzburg-Landau (GL) type current following the standard GL approach for type II superconductors \cite{ref24}-\cite{ref25}. From these computations we have observed that apart from having the usual GL piece, the local form of the current also receives some non trivial contributions that appear at the leading order in the CS coupling ($ \kappa $). As the main goal of this paper, in Section 3, we have computed the $ U(1) $ current for the boundary theory following the standard AdS/CFT prescription \cite{Hartnoll:2009sz}. In order to do that we have considered a gravitational theory in $ AdS_{4} $ that is basically coupled to the abelian Higgs model originally proposed in \cite{ref16}.  To induce an effective CS term for the boundary theory, we have added to this model a coupling of the form $ \theta F\wedge F $, where $ \theta $ is a neutral pseudo scalar field propagating in the $  AdS_{4} $ back ground\footnote{This pseudo scalar field $ \theta $ is generally known as the axion field which was observed long back as forming the non trivial scalar hair for black holes \cite{ref26}-\cite{ref27}. Theories that include axionic couplings have been further investigated under various circumstances, for example, in the context of Einstein-Maxwell Dilaton gravity \cite{ref28}-\cite{ref30}, super-gravity models \cite{ref31}-\cite{ref32}, holographic gluodynamics \cite{ref33}, super-symmetric standard model \cite{ref34}-\cite{ref35} and very recently in the context of AdS/CMT \cite{ref36}-\cite{ref37}.  In particular, the reference \cite{ref36} is quite interesting in the sense that there it has been shown for the first time that by introducing the axionic coupling to the Maxwell sector one can in fact generate Maxwell-CS like magnetic vortex solution for the boundary theory.}. The \textit{local} form of the supercurrent that has been computed using holographic techniques is found to be in remarkable agreement to that with the local form of the GL current obtained for the Maxwell-CS type vortices in ($ 2+1 $) dimensions. This eventually confirms that the coupling of the form $ \theta F\wedge F $ indeed provides an effective \textit{holographic} description for the abelian CS Higgs model in ($ 2+1 $) dimensions.  In order to check the thermodynamic stability of the vortex configuration, based on the AdS/CFT prescription, we have computed the free energy of the system in Section 4. Our analysis reveals the fact that the configuration in the symmetry broken phase possesses lesser free energy and therefore it is thermodynamically more stable. Motivated from these various evidences regarding the duality between the abelian CS Higgs model in ($ 2+1 $) dimensions and the corresponding gravity theory in $ AdS_{4} $ space time, we have explored it a bit further in Section 5, where we have explicitly computed the coherence length ($ \xi $) as well as the GL coefficient ($ \Bbbk $) \cite{Maeda:2008ir}-\cite{Zeng:2009dr} associated with Maxwell-CS type vortices in ($ 2+1 $) dimensions using the holographic techniques and obtained some corrections at the leading order in the CS coupling ($ \kappa $). Finally we have concluded in Section 6.

\section{Chern-Simons vortices in ($ 2+1 $) dimensions}
In this section, we start by reviewing the abelian CS Higgs model in ($ 2+1 $) dimensions. Our goal would be to show that using the standard GL prescription \cite{ref24}, one can in fact arrive at the GL like local expression for the current associated with the Maxwell-CS type vortices in ($ 2+1 $) dimensions. In other words, we shall follow exactly the same prescription as one does while computing the GL current for type II superconductors.  At this point it is customary to mention that through out the analysis we consider our system close to the critical point ($ T\sim T_c $) so that the scalar wave function ($ \Psi $) (which is also the \textit{order parameter} for the present case) is very small.

We start with the following Lagrangian \cite{ref39}-\cite{ref40} in ($ 2+1 $) dimensions, namely\footnote{The signature we are working with is $ \eta_{\mu\nu}=diag(-1,1,1) $.} 
\begin{eqnarray}
\mathcal{L} =-\frac{1}{4}F^{2}_{ab}-\mid D_{a}\Psi\mid^{2}+\frac{\kappa}{4}\epsilon^{abc}A_{a}F_{bc}-V(\mid \Psi \mid)\label{Eq1}
\end{eqnarray}
where, $ V(\mid \Psi \mid)= \frac{\lambda}{2}(\mid \Psi \mid^{2}-1)^{2}$ and $ D_{a}=\nabla_{a}-iq A_{a} $, with $ a=0,1,2 $.

The associated static energy functional \cite{ref39} is given by,
\begin{eqnarray}
\mathcal{E}=\int d^{2}\textbf{x} \left[\frac{1}{2m^{\ast}}\mid \left( \frac{1}{i}\nabla - q \textbf{A}\right)\Psi(\textbf{x})\mid^{2}+\frac{\lambda}{2}(\mid \Psi \mid^{2}-1)^{2}+\frac{\mathcal{H}^{2}}{8\pi}  \right]\label{E} 
\end{eqnarray}
where $ \mathcal{H} (= F_{xy}) $ is the applied magnetic field normal to the ($x-y  $) plane.

 To calculate the electromagnetic current, as a first step, we minimize the energy functional\footnote{At this stage, it is worthwhile to mention that the so called associated static energy functional ($ \mathcal{E} $) for the abelian CS Higgs theory eventually coincides with that of the usual GL free energy for type II superconductors \cite{ref39}. This is something what motivates us to compute the current for the abelian CS Higgs system by calculating the proper minima of the static energy functional. This is the standard GL prescription to compute the current associated with the vortices in type II superconductors \cite{ref24}-\cite{ref25}.} ($ \mathcal{E} $) defined above in (\ref{E}). The minimization could be done in two steps, first by considering the shift ($ \textbf{A(x)}\rightarrow \textbf{A(x)}+ \textbf{a(x)}$) in the vector potential ($ \textbf{A} $) and then setting the coefficient associated with $ \textbf{a(x)} $ equal to zero. This finally yields the current\footnote{For simplicity we have fixed all the numerical pre factors equal to identity.},
\begin{eqnarray}
\textbf{j}=\frac{1}{4\pi}\nabla \times \mathcal{H} =i \left(\Psi \nabla \Psi^{\dagger}-\Psi^{\dagger} \nabla \Psi \right)-\mid \Psi(\textbf{x})\mid^{2}\textbf{A}\label{j}. 
\end{eqnarray}

Our next goal would be to evaluate this current \textit{on-shell}, i.e; by substituting the explicit solutions for the gauge field ($ A_{\mu} $) as well as the scalar field ($ \Psi $) in to the expression for the current given above in (\ref{j}). The equations of motion could be directly read off from the Lagrangian (\ref{Eq1}) as\footnote{We have fixed the gauge as $ \nabla_aA^{a}=0 $.}, 
\begin{eqnarray}
\nabla^{2}_b A^{a}&=&-\frac{\kappa}{2}\epsilon^{abc}F_{bc}-iq\left[\Psi^{\dagger}D^{a}\Psi - \Psi (D^{a}\Psi)^{\dagger} \right]\label{F}\\
\nabla_b^{2}\Psi &=&  2iq A^{a}\nabla_a \Psi +q^{2}A_a^{2}\Psi -\lambda \Psi \label{P}.
\end{eqnarray}
The next step would be to solve the above set of equations (\ref{F}) and (\ref{P}) perturbatively in $ \kappa $ and $ \Psi $. Let us first consider the following perturbation in $ \Psi $ namely, 
\begin{eqnarray}
\Psi &=&\sqrt{\epsilon}\Psi_{1}+\mathcal{O}(\epsilon^{3/2})\nonumber\\
A_a &=& A_{a}^{(0)}+\epsilon A_{a}^{(1)}+\mathcal{O}(\epsilon^{2})\label{Eq6}
\end{eqnarray}
where\footnote{Here $ \epsilon = 1-\frac{T}{T_c} $.} $ \mid \epsilon \mid \ll 1 $. Here $ \Psi_{1} $ stands for the first non trivial fluctuation in the scalar profile ($ \Psi $). The various \textit{superscripts} in the gauge field components correspond to different order of fluctuations in the gauge fields due to the presence of the non trivial scalar profile, for example, $ A_{a}^{(0)} $s are the solutions of (\ref{F}) corresponding to $ \Psi =0 $ namely,
\begin{eqnarray}
\nabla^{2}_b A^{a(0)}=-\frac{\kappa}{2}\epsilon^{abc}F^{(0)}_{bc}.\label{Maxwell}
\end{eqnarray}

Let us now solve the above equation (\ref{Maxwell}) perturbatively in $ \kappa $. Consider the following perturbative expansion,
\begin{eqnarray}
A_a^{(0)} = A_{a}^{(0)(\kappa^{(0)})}+\kappa A_{a}^{(0)(\kappa^{(1)})}+\mathcal{O}(\kappa^{2})\label{Eq8}
\end{eqnarray}
where $A_{a}^{(0)(\kappa^{(0)})}$ corresponds to the solution of (\ref{Maxwell}) while $ \kappa =0 $, $A_{a}^{(0)(\kappa^{(1)})}$  stands for the first non-trivial leading order correction to $ A_{a}^{(0)} $ due to CS coupling ($ \kappa $) and so on. Our first step would be to solve $ A_{a}^{(0)(\kappa^{(0)})} $ from (\ref{Maxwell}). In order to do that we take the following ansatz,
\begin{eqnarray}
A_{a}^{(0)(\kappa^{(0)})}=(A_{t}^{(0)(\kappa^{(0)})}(\textbf{x}),~0,~A_{y}^{(0)(\kappa^{(0)})}(x))
\end{eqnarray}
where the individual components satisfy equations of the following type\footnote{Here $ \Delta = \partial_x^{2}+\partial_y^{2} $ is the usual Laplacian.},
\begin{eqnarray}
\Delta A_{t}^{(0)(\kappa^{(0)})} &=& 0\nonumber\\
\partial_x^{2}A_{y}^{(0)(\kappa^{(0)})}&=& 0.
\end{eqnarray}

The first equation is nothing but the La place's equation in two dimensions the solution of which should vanish at large values of spatial coordinates. The above equations suggest that we may take $ A_{t}^{(0)(\kappa^{(0)})}\sim e^{-\varrho x}e^{i\varrho y} $ and $ A_{y}^{(0)(\kappa^{(0)})}= \mathcal{H}_{c2}x $, where $ \varrho $ is some constant that does not depend on the spatial coordinates ($ \textbf{x} $) and $ \mathcal{H}_{c2} $ is the upper value of the critical magnetic field above which the condensate vanishes. At this point one should take a note on the fact that as one moves away from the origin ($\sqrt{\mid \textbf{x}\mid} \gg\frac{1}{\varrho} $), the temporal part of the gauge field tends to vanish i.e; $A_{t}^{(0)(\kappa^{(0)})} \approx 0 $. As a consequence of this we shall see later on that in order to appreciate the effect of CS coupling on the expression of the current one can not move arbitrarily far away from the vortex. 

Let us now write down (\ref{Maxwell}) considering the leading order effect in the CS coupling ($ \kappa $). The corresponding equations could be immediately read off as\footnote{Here $ i(=x,y) $ corresponds to spatial coordinates.},  
\begin{eqnarray}
\Delta A_{t}^{(0)(\kappa^{(1)})} &=& \mathcal{H}_{c2}\\
\Delta A_{i}^{(0)(\kappa^{(1)})}&=& -  \epsilon_{i}^{j}\partial_j A_{t}^{(0)(\kappa^{(0)})}\label{A}
\end{eqnarray}
where we have used the notation $ \epsilon^{tij}=\epsilon^{ij}=-\epsilon^{ji} $. We will be actually interested to note down the solution corresponding to the last equation (\ref{A}) which turns out to be\footnote{ We have used the fact $  \partial_{j}'G(\textbf{x}-\textbf{x}')=- \partial_{j}G(\textbf{x}-\textbf{x}') $.},
\begin{eqnarray}
A_{i}^{(0)(\kappa^{(1)})}=  \epsilon_{i}^{j}\partial_j\int d^{2}\textbf{x}' A_{t}^{(0)(\kappa^{(0)})}(\textbf{x}')G(\textbf{x}-\textbf{x}')=\epsilon_{i}^{j}\partial_j\Pi (\textbf{x})\label{Eq13}
\end{eqnarray}
where $ G(\textbf{x}-\textbf{x}') $ is the Green's function on the two dimensional plane\footnote{Two dimensional Green's function satisfies the equation $ \Delta G(\textbf{x}-\textbf{x}')=-\delta(\textbf{x}-\textbf{x}') $.}.

Let us now turn on the equation corresponding to the leading order fluctuation in the scalar field namely $\Psi= \Psi_{1} $. Since the gauge fields ($ A^{(0)}_a $) corresponding to $ \Psi=0 $ themselves contain the explicit $ \kappa $ corrections (see Eq.(\ref{Eq8})), therefore from (\ref{P}) we can also solve $ \Psi_1 $ perturbatively in $ \kappa $. This inspires us to write down the following expansion for $ \Psi_{1} $ namely,
\begin{eqnarray}
\Psi_{1} = \Psi^{(0)}_{1}+\kappa \Psi^{(1)}_{1}+\mathcal{O}(\kappa^{2})\label{Eq14}
\end{eqnarray}
where $ \Psi^{(0)}_{1} $ corresponds to the solution of (\ref{P}) for $ \kappa=0 $, $ \Psi^{(1)}_{1} $ is the first non trivial $ \kappa $ correction to $ \Psi_{1} $ and so on. Substituting (\ref{Eq6}) and (\ref{Eq14}) in to (\ref{j}), one can express the (GL) current in the following series as,
\begin{eqnarray}
j_{i} = j_{i}^{(0)}+j_{i}^{(1)}+ ..\label{Eq15}
\end{eqnarray}
where each individual term reads as\footnote{Here we have rescaled each individual $ j_{i} $s by the factor $ \epsilon $.},
\begin{eqnarray}
j_{i}^{(0)}&=&i \left( \Psi_{1}^{(0)} \nabla_{i} \Psi_{1}^{(0)\dagger}-\Psi_{1}^{(0)\dagger} \nabla_{i} \Psi_{1}^{(0)}\right) -\mid \Psi_{1}^{(0)}(\textbf{x})\mid^{2}A^{(0)}_{i}\nonumber\\
j_{i}^{(1)}&=& \kappa\left[ i \left( \Psi_{1}^{(0)} \nabla_{i} \Psi_{1}^{(1)\dagger}+\Psi_{1}^{(1)} \nabla_{i} \Psi_{1}^{(0)\dagger}-\Psi_{1}^{(0)\dagger} \nabla_{i} \Psi_{1}^{(1)}-\Psi_{1}^{(1)\dagger} \nabla_{i} \Psi_{1}^{(0)}\right)-\left(\Psi_{1}^{(0)\dagger}\Psi_{1}^{(1)}+\Psi_{1}^{(1)\dagger}\Psi_{1}^{(0)}\right)A^{(0)}_{i} \right]\nonumber\\
..&..&~~..~~..\label{Eq16}  
\end{eqnarray}
Notice that here we have evaluated the current only considering the first non trivial fluctuation in the scalar field, namely $ \Psi =\Psi_{1} $. One should also take note on the fact that, if for example, we did not have any CS term in our action from the very beginning i.e; we would have started with only the usual abelian Higgs model, then of course all the currents like $ j_{i}^{(1)} $, $ j_{i}^{(2)} $ would have been disappeared from our theory because these are the pieces that are always coupled with various powers of the CS coupling\footnote{In other words, in order to compute these currents one needs to know the scalar fluctuations appearing at various orders in the CS coupling namely $ \Psi_{1}^{(1)} $, $ \Psi_{2}^{(2)} $ etc.} ($ \kappa $). On the other hand, the first term in the expansion of (\ref{Eq15}), namely  $ j_{i}^{(0)} $ has two parts in it - one that is independent of the CS effect\footnote{The quantity $ \Psi_{1}^{(0)} $ does not contain any information about the CS coupling. Therefore this could be regarded as the effect that is appearing solely from the abelian Higgs sector.} and the other part that contains explicit $ \kappa $ corrections in various orders since according to (\ref{Eq8}) one can have a perturbative expansion for the gauge field $ A^{(0)}_{i} $ in powers of $ \kappa $. This enforces us to conclude that the current $ j_{i}^{(0)} $ receives contribution both from the usual Maxwell sector as well as from the CS sector and therefore it is the current associated with \textit{mixed} Maxwell-CS type vortices\footnote{In the present analysis our focus would be to calculate this \textit{mixed} current $ j_{i}^{(0)} $ using a holographic set up. This will be sufficient to capture the effective physics of the abelian CS Higgs model because all the subsequent terms like $ j_{i}^{(m)} $ (for $ m\geq 1 $  ) in the expansion (\ref{Eq15}) are small compared to the leading term since they appear with higher powers in $ \kappa $ and also contain higher order fluctuations.}.

Before we proceed further, let us first note that $ \Psi_{1}^{(0)}  $ and $ \Psi_{1}^{(1)}  $ satisfy equations of the following type\footnote{In the first equality we have essentially considered a small region around the core of the vortex namely  $ \sqrt{\mid \textbf{x} \mid} \ll \frac{1}{\varrho} $ so that the radius of convergence of that region $ R\ll\frac{1}{\varrho} $. Therefore the \textit{local} expression for the current that we would obtain here is valid around a closed neighbourhood of the vortex whose area is $ 2 \pi R^{2} $.  Also note that here we have defined $ \tilde{\lambda}=\lambda +q^{2} $.},
\begin{eqnarray}
(\Delta - 2iq\mathcal{H}_{c2}x\partial_{y}-q^{2}\mathcal{H}_{c2}^{2}x^{2})\Psi_{1}^{(0)}-\tilde{\lambda}\Psi_{1}^{(0)}= 0\label{H}
\end{eqnarray}
and,
\begin{eqnarray}
\left[ \nabla_{a}^{2}-2iqA^{a(0)(\kappa^{(0)})}\nabla_a -\left( q^{2}A_{a}^{2(0)(\kappa^{(0)})}+\lambda\right)\right]  \Psi_{1}^{(1)} =2iqA^{a(0)(\kappa^{(1)})}\nabla_a \Psi_{1}^{(0)} \nonumber\\+2q^{2}A^{a(0)(\kappa^{(0)})}A_{a}^{(0)(\kappa^{(1)})}\Psi_{1}^{(0)}.
\end{eqnarray}
Since our primary goal is to evaluate the \textit{mixed} current $ j_{i}^{(0)} $, therefore we will be primarily interested in solving (\ref{H}). From (\ref{H}) we note that there exists a non trivial potential along the $ x $ direction while the motion along the $ y $ direction does not seem to have any such constraints. Based on these observations we take the following ansatz for the scalar field $ \Psi_{1}^{(0)} $ namely,
\begin{eqnarray}
\Psi_{1}^{(0)}=e^{ip_{y}y}X(x)\label{Eq19}.
\end{eqnarray}
Substituting (\ref{Eq19}) in to (\ref{H}) we find\footnote{Here $ \xi^{2}=\frac{1}{\tilde{\lambda}} $.},
\begin{eqnarray}
 -X^{''}(x)+q^{2}\mathcal{H}_{c2}^{2}\left( x-\frac{p_{y}}{q\mathcal{H}_{c2}}\right)^{2}X(x)=\frac{X(x)}{\xi^{2}}\label{Eq20}.
 \end{eqnarray}
 A number of comments are to made at this stage regarding (\ref{Eq20}). First of all (\ref{Eq20}) depicts exactly the equation for an one dimensional harmonic oscillator whose equilibrium position has been shifted by an amount $ x_{0}=(\frac{p_{y}}{q\mathcal{H}_{c2}}) $.
 In the standard GL theory for type II superconductors one encounters exactly the same equation where one can identify $ \xi^{2}(=\frac{1}{\tilde{\lambda}}) $ as the coherence length\footnote{In Section 5, we shall explicitly derive this coherence length ($ \xi $) using the holographic techniques.} pertaining to the system \cite{ref24}-\cite{ref25} near the critical point ($ T=T_c $) of the phase transition line. Here $ p_y $ is associated with the periodicity along $ y $ direction,
 \begin{eqnarray}
 p_{y}=\frac{2\pi l}{a_{y}},~~~l\in Z
 \end{eqnarray}
 where the coefficient $ a_y $ stands for the periodicity along $ y $ direction.
 
 The most general lowest energy ($ n=0 $) solution could be expressed as a linear superposition of a set of (eigen) functions corresponding to different values of $ l $ namely,
\begin{eqnarray}
\Psi_{1}^{(0)} = \sum_{l=-\infty}^{l=\infty}c_{l}exp\left(\frac{2i\pi yl }{a_{y}} \right)exp\left(-\frac{1}{2\xi^{2}}\left( x-\frac{2\pi l\xi^{2}}{a_{y}}\right)^{2}  \right)  
\end{eqnarray}
where the coefficients $ c_l $ could be expressed as,
\begin{eqnarray}
c_l = exp\left( -\frac{i \pi a_x \xi^{2}l^{2}}{a_y^{2}}\right). 
\end{eqnarray} 

Using the \textit{elliptic theta} function one can express the most general solution for (\ref{H}) as \cite{Maeda:2009vf},
\begin{eqnarray}
\Psi_{1}^{(0)}=e^{-\frac{x^{2}}{2\xi^{2}}}\vartheta_{3}(v,\tau)\label{Eq21}
\end{eqnarray}
where the \textit{elliptic theta} function could be formally expressed as\footnote{Here $ a_x $ and $ a_y $ are two arbitrary parameters. The parameter $ a_y $ is particularly associated with the periodicity along the $ y $ direction.},
\begin{eqnarray}
\vartheta_{3}(v,\tau)=\sum_{l=-\infty}^{l=\infty}q^{l^{2}}z^{2l}\label{Eq22}
\end{eqnarray}
 with,
 \begin{eqnarray}
 q&=&exp(i\pi \tau)= exp\left( i\pi\xi^{2}\frac{2\pi i - a_x}{a_y^{2}}\right) \nonumber\\
 z&=&exp(i\pi v)=exp\left( i\pi \frac{y-ix}{a_y}\right).
 \end{eqnarray}
Finally substituting (\ref{Eq21}) in to the expression for $ j_{i}^{(0)} $ (Eq.(\ref{Eq16})) we find\footnote{The details of the calculation have been provided in the Appendix A.}, 
\begin{eqnarray}
j^{(0)}_{i}\sim -\epsilon_{i}^{j}\partial_{j}\mid \Psi^{(0)}_{1}(\textbf{x})\mid^{2}+\kappa \Pi(\Delta)\epsilon_{i}^{j}\partial_{j}\mid \Psi_{1}^{(0)}(\textbf{x})\mid^{2}+\mathcal{O}(\kappa^{2})\label{Eq24}
\end{eqnarray}
where, $\Pi(\Delta)=\int_{-\frac{\Delta}{2}}^{\frac{\Delta}{2}}d^{2}\textbf{x}~~ln\mid\textbf{x}\mid $. Here $ \Delta $ could be regarded as an effective scale (or, the width of the curve) that essentially measures how rapidly the two point correlation (in two dimensions) dies off as one moves away from the origin.  

Eq.(\ref{Eq24}) expresses the current as a \textit{local} function of the vortex solution in a small neighbourhood close to the centre of the vortex. Note that here the first term in (\ref{Eq24}) is exactly the piece that appears in the usual GL theory for type II superconductors  \cite{ref24}. The rest is the leading order correction to this current due to the CS term. Therefore the current $  j^{(0)}_{i}$ receives a mixed contribution from both the Maxwell as well as the CS sector. The goal of our computation in the next Section would be to reproduce this result based on the standard $AdS_{4}/CFT_{3}$ prescription.

 \section{Supercurrent: $ AdS_{4}$/$CFT_{3}$ correspondence}
  
\subsection{The bulk theory}
In this part of our analysis, we consider a gravitational theory (defined over an $ AdS_{4} $ background) that is coupled to the abelian Higgs model originally proposed in \cite{ref16}. In addition to that, to produce an effective CS term for the boundary theory we add to the matter content of the Lagrangian a term of the form $ \theta F\wedge F $. Combining all these facts, we consider the following action as the starting point of our analysis,
\begin{eqnarray}
S=\frac{1}{16 \pi G_{4}}\int d^{4}x \sqrt{-g}\left[ R- 2\Lambda + \mathcal{L}_{m}\right] \label{Eq25}
\end{eqnarray}
where,
\begin{eqnarray}
\mathcal{L}_{m}=  - \frac{1}{4}F^{\mu\nu}F_{\mu\nu}-|\nabla_{\mu}\Psi -iA_{\mu}\Psi|^{2}-m^{2}|\Psi|^{2} - \frac{\kappa}{\sqrt{-g}}\theta \epsilon^{\mu\nu\rho\sigma}F_{\mu\nu}F_{\rho\sigma}-(\nabla_{\mu}\theta)^{2}\label{Eq26}
\end{eqnarray}
with $ \Lambda( = -\frac{3}{l^{2}}) $ as the cosmological constant\footnote{In the subsequent analysis we shall set $ l=1 $.}. The first three terms on the r.h.s. of (\ref{Eq26}) correspond to the standard abelian Higgs model \cite{ref16} which consists of a $ U(1) $ gauge field minimally coupled to the complex scalar field ($ \Psi $). Here $ \theta $ is a massless neutral \textit{pseudo} scalar field. The coupling ($ \kappa $) associated with $ \theta F\wedge F $ term serves as the effective CS coupling for the boundary theory\footnote{Since $ \theta $ eventually tends towards a constant value near the boundary, therefore actually it is the coefficient ($ \kappa \theta $) sitting in front of the topological term $ F\wedge F $ that serves as the effective CS coupling for the boundary theory\cite{ref36}.}\cite{ref36}. 

Considering the \textit{probe} limit\footnote{The probe limit essentially corresponds to the fact that the matter fields do not back react on the back ground space time \cite{ref17}. This is essentially achieved by rescaling $ A_{\mu}\rightarrow \frac{A_{\mu}}{q} $ and $ \Psi \rightarrow \frac{\Psi}{q} $ and then taking the large $ q $ limit while keeping the numerator fixed. }, we perform the entire analyses over the fixed background of an asymptotically $ AdS_4 $ black brane solution namely, 
  \begin{eqnarray}
 ds^{2}=-f(u)dt^{2}+\frac{r_+^{2}}{u^{4}}f^{-1}(u)du^{2}+\frac{r_+^{2}}{u^{2}}d\textbf{x}^{2}
 \end{eqnarray}
 where,
 \begin{eqnarray}
 f(u)=\frac{r_+^{2}}{u^{2}}(1-u^{3}).
 \end{eqnarray}
 
 Note that with the above choice of coordinates the horizon is located at $ u=1 $ whereas the boundary of the $ AdS_{4} $ space time is located at $ u=0 $. The temperature of the black brane is given by, 
  \begin{eqnarray}
 T=\frac{3r_{+}}{4\pi}\label{eq5}
 \end{eqnarray}
 which we consider to be fixed for the present analyses. As a consequence of this one may consider the boundary theory to be at the same temperature as that of the black brane.
 
Considering (\ref{Eq25}) as the starting point of our analysis, our first step would be to study the dynamics of the matter fields in the bulk $ AdS_{4} $ space time. In order to do that, we first write down the following equations of motion namely,
\begin{eqnarray}
\nabla_{\mu}F^{\nu\mu}+\frac{4\kappa}{\sqrt{-g}}\epsilon^{\nu\mu\rho\sigma}\nabla_{\mu}\theta F_{\rho\sigma}&=&j^{\nu}\nonumber\\
\nabla^{\mu}\nabla_{\mu}\Psi -iA^{\mu}\nabla_{\mu}\Psi - i\nabla_{\mu}(A^{\mu}\Psi)-A_{\mu}^{2}\Psi - m^{2}\Psi &=& 0\nonumber\\
\nabla_{\mu}^{2}\theta -\frac{\kappa}{2\sqrt{-g}}\epsilon^{\mu\nu\rho\sigma}F_{\mu\nu}F_{\rho\sigma}&=&0\label{Eq30}
\end{eqnarray} 
 where,
 \begin{eqnarray}
 j^{\nu}=i(\Psi(D^{\nu}\Psi)^{\dagger}-\Psi^{\dagger}D^{\nu}\Psi).
  \end{eqnarray} 

 As a next step we solve these above set of equations (\ref{Eq30}) with the following perturbative technique. Let us consider the following perturbative expansion of the form\footnote{Here $ \varepsilon_{H}(=\frac{\mathcal{H}_{c2}-H}{\mathcal{H}_{c2}}) $ is the perturbation parameter such that $ \mid\varepsilon_{H}\mid \ll 1 $, and $ \mathcal{H}_{c2} $ is the upper critical magnetic field.},
 \begin{eqnarray}
 A_{\mu}&=& A_{\mu}^{(0)}+\varepsilon_{H} A_{\mu}^{(1)}(u,\textbf{x})+\mathcal{O}(\varepsilon_{H}^{2})\nonumber\\
 \theta &=& \theta^{(0)}+\varepsilon_{H} \theta^{(1)}(u,\textbf{x})+\mathcal{O}(\varepsilon_{H}^{2})\nonumber\\
 \Psi &=& \varepsilon_H^{1/2}\psi_{1}(u,\textbf{x})+\mathcal{O}(\varepsilon_{H}^{3/2}).\label{Eq32}
 \end{eqnarray}

 The above set of expansions automatically reflect the fact that the scalar field ($ \Psi $) acts as the order parameter for our theory which therefore vanishes at the (upper) critical value of the magnetic field $ H=\mathcal{H}_{c2} $ \cite{Maeda:2009vf}-\cite{Banerjee:2013maa}. The superscript ($ m $) for the gauge fields ($ A_{\mu}^{(m)} $) as well as the pseudo scalar field ($ \theta^{(m)} $) corresponds to fluctuations at different order due to the presence of the non trivial scalar profile ($ \Psi $), for example, $ m=0 $ stands for the solutions corresponding to $ \Psi=0 $, i.e; when we are exactly at the critical point. On the other hand, $ m=1 $ corresponds to the first non trivial fluctuations corresponding to $ \Psi=\psi_{1} $ and so on. Our next goal would be to solve these fluctuations explicitly upto leading order.
 
 \subsubsection{Zeroth order equations}
Let us first consider the equations at the zeroth order level. Substituting (\ref{Eq32}) in to (\ref{Eq30}) we find the following set of equations namely, 
\begin{eqnarray}
\nabla_{\mu}F^{(0)\nu\mu}+\frac{4\kappa}{\sqrt{-g}}\epsilon^{\nu\mu\rho\sigma}\nabla_{\mu}\theta^{(0)} F^{(0)}_{\rho\sigma}&=&0\nonumber\\
\nabla_{\mu}^{2}\theta^{(0)} -\frac{\kappa}{2\sqrt{-g}}\epsilon^{\mu\nu\rho\sigma}F^{(0)}_{\mu\nu}F^{(0)}_{\rho\sigma}&=&0.\label{Eq33}
\end{eqnarray} 
In order to solve (\ref{Eq33}) we take the following ansatz,
\begin{eqnarray}
A^{(0)}_{\mu}&=&(\varphi(u),0,0,\mathcal{H}_{c2}x)\nonumber\\
\theta^{(0)}&=&\tau (u).\label{eq10}
\end{eqnarray} 
Substituting (\ref{eq10}) in to (\ref{Eq33}) we find,
\begin{eqnarray}
\varphi''(u)+\frac{8\kappa\mathcal{H}_{c2}}{r_+}\tau'(u)&=&0\nonumber\\
\tau''(u)+\frac{f'(u)}{f(u)}\tau'(u)+\frac{4\kappa\mathcal{H}_{c2}}{r_+f(u)}\varphi'(u)&=&0.\label{Eq35}
\end{eqnarray} 
Our next goal would be to solve these equations perturbatively in $ \kappa $ near the boundary ($ u\rightarrow 0 $) of the $ AdS_{4} $. In order to do that we solve these equations as a perturbation in the CS coupling ($ \kappa $). Consider the following perturbative expansion,
\begin{eqnarray}
\varphi &=& \varphi^{(\kappa^{(0)})}+\kappa \varphi^{(\kappa^{(1)})}+\mathcal{O}(\kappa^{2})\nonumber\\
\tau &=& \tau^{(\kappa^{(0)})}+\kappa \tau^{(\kappa^{(1)})}+\mathcal{O}(\kappa^{2})\label{Eq36}
\end{eqnarray} 
where the superscript $ \kappa^{(0)} $ corresponds to the solutions of (\ref{Eq35}) for $ \kappa=0 $. On the other hand, $\varphi^{(\kappa^{(1)})}  $ (or $ \tau^{(\kappa^{(1)})} $) stands for the leading order correction to $ \varphi^{(\kappa^{(0)})} $ (or $ \tau^{(\kappa^{(0)})} $ ).

 Substituting (\ref{Eq36}) in to (\ref{Eq35}) we arrive at the following set of solutions which could be enumerated as, 
 \begin{eqnarray}
 \varphi^{(\kappa^{(0)})}&=&\mu (1-u)\nonumber\\
 \varphi^{(\kappa^{(1)})}&=&\mu \left( 1+\frac{2\mathcal{C}_{2}\mathcal{H}_{c2}}{3\mu r_+^{3}}\right)(1-u)+\mathcal{O}(u^{4})\nonumber\\
 \tau^{(\kappa^{(0)})}&=& \tau^{(\kappa^{(1)})}= \mathcal{C}_{1}+\frac{\mathcal{C}_{2}}{3r_+^{2}}u^{3}+\mathcal{O}(u^{6})\label{theta} 
 \end{eqnarray}
 where $ \mu $ is the chemical potential of the boundary theory and $ \mathcal{C}_{1} $(or $ \mathcal{C}_{2} $) is some arbitrary constant. At this point one might take a note on the fact that the pseudo scalar field has a faster fall off near the boundary of the $ AdS_{4} $ as compared to that of the $ U(1) $ gauge field and eventually becomes a constant ($ \sim \mathcal{C}_{1} $) at the boundary. Therefore at the boundary of the $ AdS_{4} $ we have an effective CS coupling of the form $ \kappa \mathcal{C}_{1}$ \cite{ref36}.

 \subsubsection{Equation for scalar field}
Let us now consider the scalar field equation for the leading order fluctuations, namely $ \Psi=\psi_{1} $ which reads as, 
\begin{eqnarray}
 \partial_{u}^{2}\psi_{1}+\frac{f^{'}(u)}{f(u)}\partial_{u}\psi_{1}+\frac{r_{+}^{2}\varphi^{2}}{u^{4}f^{2}(u)}\psi_{1}-\frac{m^{2}r_{+}^{2}}{u^{4}f(u)}\psi_{1}+ \frac{1}{u^{2}f(u)}(\Delta -2i\mathcal{H}_{c2}x\partial_{y}-\mathcal{H}_{c2}^{2}x^{2})\psi_{1}=0.\label{Eq38}
 \end{eqnarray} 
 
 It is quite intuitive to solve the above equation (\ref{Eq38}) using the method of separation of variables,
 \begin{eqnarray}
 \psi_{1}(u,\textbf{x})=\Re(u)e^{ik_{y}y}X(x)=\Re(u)\mathcal{X}(\textbf{x}).
 \end{eqnarray}
 On substitution in to (\ref{Eq38}), we arrive at the following set equations namely,
  \begin{eqnarray}
 \Re'' (u) +\frac{f^{'}(u)}{f(u)}\Re' (u) +\frac{r_{+}^{2}\varphi^{2}}{u^{4}f^{2}(u)}\Re(u)-\frac{m^{2}r_{+}^{2}}{u^{4}f(u)}\Re(u)=\frac{\Re(u)}{\xi^{2}u^{2}f(u)}\label{Eq40}
 \end{eqnarray}
 and,
 \begin{eqnarray}
 -X^{''}(x)+\mathcal{H}_{c2}^{2}\left( x-\frac{k_{y}}{\mathcal{H}_{c2}}\right)^{2}X(x)=\frac{X(x)}{\xi^{2}}.\label{Eq41}
 \end{eqnarray}
 It is quite interesting to note that (\ref{Eq41}) has exactly the same structure as that of the (\ref{Eq20}) which we have obtained earlier in the context of Maxwell-CS vortices where we did not have any gravity at all. The remarkable fact here is that we are getting the same equation using a gravity model. 
 Like in the previous case, here $ k_y $ is associated with the periodicity along $ y $ direction as,
 \begin{eqnarray}
 k_{y}=\frac{2\pi l}{a_{y}},~~~l\in Z
 \end{eqnarray}
 where the coefficient $ a_y $ is associated with the periodicity along $ y $ direction.
 
Finally, using the \textit{elliptic theta} function (\ref{Eq22}), the above solution could be rewritten as \cite{Maeda:2009vf},
\begin{eqnarray}
\mathcal{X}(\textbf{x})=e^{-\frac{x^{2}}{2\xi^{2}}}\vartheta_{3}(v,\tau)\label{Eq45}
\end{eqnarray} 
where $ \xi $ is again playing the role of the coherence length. 

One important remark that has to be made at this stage is the following: From (\ref{Eq45}) we note that the vortex solution has an exponential die off along the $ x $ direction and it eventually vanishes for $ \mid x \mid \gg \xi $. This implies that the vortex structure has a natural length scale which is roughly of the order of the coherence length ($ \xi $) itself.


\subsubsection{Solving the fluctuations}
 Let us now consider the gauge field as well as the equation for the pseudo scalar field ($ \theta $) corresponding to the fluctuations at the leading order. These equations turn out to be the following,
 \begin{eqnarray}
\nabla_{\mu}F^{(1)\nu\mu}+\frac{4\kappa}{\sqrt{-g}}\epsilon^{\nu\mu\rho\sigma}(\nabla_{\mu}\theta^{(0)} F^{(1)}_{\rho\sigma}+\nabla_{\mu}\theta^{(1)} F^{(0)}_{\rho\sigma})&=&j^{(1)\nu}\nonumber\\
\nabla_{\mu}^{2}\theta^{(1)} -\frac{\kappa}{\sqrt{-g}}\epsilon^{\mu\nu\rho\sigma}F^{(0)}_{\mu\nu}F^{(1)}_{\rho\sigma}&=&0.\label{Eq48}
\end{eqnarray} 
Let us first write down the above set of equations (\ref{Eq48}) explicitly in to different components namely\footnote{Here $ \sigma (\textbf{x})(=\mid \mathcal{X}(\textbf{x})\mid^{2} ) $ corresponds to the total number of particles forming the charge condensate per unit volume around the point $ \textbf{x} $. },
\begin{eqnarray}
\mathcal{D}_t A^{(1)}_{t}+\frac{8\kappa u^{2}f(u)}{r_+}\left[\tau'(u)F^{(1)}_{xy}(u,\textbf{x})+\mathcal{H}_{c2}\partial_u \theta^{(1)}(u,\textbf{x}) \right]&=&\frac{2r_+^{2}}{u^{2}}\varphi(u)\Re^{2}(u)\sigma (\textbf{x})\nonumber\\
\mathcal{D}_{s}A^{(1)}_{x}-8\kappa r_+\left[\tau'(u)\partial_y A^{(1)}_{t}(u,\textbf{x})-\varphi'(u)\partial_y \theta^{(1)}(u,\textbf{x}) \right]&=&\frac{r_+^{2}}{u^{2}}\Re^{2}(u)\epsilon_{x}^{y}\partial_y\sigma (\textbf{x})\nonumber\\
\mathcal{D}_{s}A^{(1)}_{y}-8\kappa r_+\left[\varphi'(u)\partial_x \theta^{(1)}(u,\textbf{x})-\tau'(u)\partial_x A^{(1)}_{t}(u,\textbf{x})\right]&=&\frac{r_+^{2}}{u^{2}}\Re^{2}(u)\epsilon_{y}^{x}\partial_x\sigma (\textbf{x})\nonumber\\ 
\mathcal{D}_{\theta}\theta^{(1)}+\frac{4\kappa u^{2}}{r_+}\left[ \mathcal{H}_{c2}\partial_u A^{(1)}_{t}(u,\textbf{x})+\varphi'(u)F^{(1)}_{xy}(u,\textbf{x})\right]&=&0\label{Eq47} 
\end{eqnarray} 
 where the various differential operators have the following form,
 \begin{eqnarray}
 \mathcal{D}_t &=& u^{2}f(u)\partial_{u}^{2}+\Delta\nonumber\\
 \mathcal{D}_s &=& \partial_u (u^{2}f(u)\partial_{u})+\Delta\nonumber\\
 \mathcal{D}_{\theta}&=& u^{2}\partial_u (f(u)\partial_u)+\Delta.
 \end{eqnarray}
 
 At this stage it is important to note that in order to arrive at (\ref{Eq47}) we have used two crucial facts, firstly, we have fixed the gauge $ A^{(1)}_u=0 $ at the leading order in the fluctuations and secondly using the residual gauge symmetry $ A^{(1)}_{i}\rightarrow A^{(1)}_{i} - \partial_i \varpi (\textbf{x}) $ at the level of equations of motion, we have made a further gauge choice namely, $ \partial_i A^{(1)}_{i} =0 $.
 
 The equations above in (\ref{Eq47}) correspond to a set of non liner coupled second order differential equations. In general one can write down a solution for such equations in terms of the Green's function. Following the same prescription and considering whatever remaining on the r.h.s. as the source term, we finally express these fluctuations in terms of integral over some suitable Green's function that satisfy certain $ AdS_{4} $ boundary conditions.
  
Our next step is quite intuitive i.e; we would first write down (\ref{Eq47}) perturbatively in the CS coupling ($ \kappa $). Let us consider the following perturbative expansion\footnote{Considering (\ref{Eq36}) and (\ref{Eq40}) it is quite intuitive to note that the radial function $ \Re (u) $ has a natural perturbative expansion in the CS coupling ($ \kappa $).},
\begin{eqnarray}
A^{(1)}_{\mu}&=& \mathcal{A}^{(1)(\kappa^{(0)})}_{\mu}+\kappa \mathcal{A}^{(1)(\kappa^{(1)})}_{\mu} + \mathcal{O}(\kappa^{2})\nonumber\\
\theta^{(1)}&=&\theta^{(1)(\kappa^{(0)})}+\kappa \theta^{(1)(\kappa^{(1)})}+\mathcal{O}(\kappa^{2})\nonumber\\
\Re (u)&=& \Re^{(\kappa^{(0)})}+\kappa \Re^{(\kappa^{(1)})} + \mathcal{O}(\kappa^{2}).\label{Eq49}
\end{eqnarray}

Substituting (\ref{Eq49}) in to (\ref{Eq47}) we eventually arrive at various sets of equations corresponding to different order in the CS coupling ($ \kappa $). Let us first consider the equations corresponding to zeroth order in the CS coupling ($ \kappa $) which turns out to be, 
 \begin{eqnarray}
\mathcal{D}_t \mathcal{A}^{(1)(\kappa^{(0)})}_{t} &=&\frac{2r_+^{2}}{u^{2}}\varphi^{(\kappa^{(0)})}(u)\Re^{2(\kappa^{(0)})}(u)\sigma (\textbf{x})\nonumber\\
\mathcal{D}_{s}\mathcal{A}^{(1)(\kappa^{(0)})}_{x}&=&\frac{r_+^{2}}{u^{2}}\Re^{2(\kappa^{(0)})}(u)\epsilon_{x}^{y}\partial_y\sigma (\textbf{x})\nonumber\\
\mathcal{D}_{s}\mathcal{A}^{(1)(\kappa^{(0)})}_{y}&=&\frac{r_+^{2}}{u^{2}}\Re^{2(\kappa^{(0)})}(u)\epsilon_{y}^{x}\partial_x\sigma (\textbf{x})\nonumber\\ 
\mathcal{D}_{\theta}\theta^{(1)(\kappa^{(0)})}&=&0.\label{Eq50}
\end{eqnarray}

The solutions corresponding to the first three equations in (\ref{Eq50}) could be expressed as,
\begin{eqnarray}
\mathcal{A}^{(1)(\kappa^{(0)})}_{t}&=& - 2r_+^{2} \int_{0}^{1}du' \frac{\varphi^{(\kappa^{(0)})}(u')}{u'^{2}}\Re^{2(\kappa^{(0)})}(u')\int d\textbf{x}' \mathcal{G}_{t}(u,u';\textbf{x},\textbf{x}')\sigma (\textbf{x}')\nonumber\\
\mathcal{A}^{(1)(\kappa^{(0)})}_{i}&=& a_{i}(\textbf{x})- r_+^{2}\epsilon_{i}^{j} \int_{0}^{1} \frac{du'}{u'^{2}}\Re^{2(\kappa^{(0)})}(u')\int d\textbf{x}' \mathcal{G}_{s}(u,u';\textbf{x},\textbf{x}')\partial_j \sigma (\textbf{x}')\label{Eq51}
\end{eqnarray}
where $ \mathcal{G}_{t}(u,u';\textbf{x},\textbf{x}') $ and $ \mathcal{G}_{s}(u,u';\textbf{x},\textbf{x}') $ are the Green's functions which satisfy the following differential equations namely,
\begin{eqnarray}
\mathcal{D}_t  \mathcal{G}_{t}(u,u';\textbf{x},\textbf{x}') &=& - \delta (u-u')\delta (\textbf{x}-\textbf{x}')\nonumber\\
\mathcal{D}_s  \mathcal{G}_{s}(u,u';\textbf{x},\textbf{x}') &=& - \delta (u-u')\delta (\textbf{x}-\textbf{x}')
\end{eqnarray}
along with the following AdS boundary conditions,
\begin{eqnarray}
\mathcal{G}_t(u,u^{'};\textbf{x},\textbf{x}')|_{u=0}=\mathcal{G}_t(u,u^{'};\textbf{x},\textbf{x}^{'})|_{u=1}=0\nonumber\\
\mathcal{G}_{s}(u,u^{'};\textbf{x},\textbf{x}')|_{u=0}=u^{2}f(u)\partial_u\mathcal{G}_{s}(u,u^{'};\textbf{x},\textbf{x}^{'})|_{u=1}=0.
\end{eqnarray}

In order to solve the last equation in (\ref{Eq50}), we consider the following functional form,
\begin{eqnarray}
\theta^{(1)(\kappa^{(0)})}(u,\textbf{x})= \mathcal{U}(u)\mathcal{W}(\textbf{x}).\label{Eq54}
\end{eqnarray} 
Substituting (\ref{Eq54}) in to the last equation of (\ref{Eq50}) we arrive at the following set of equations namely,
\begin{eqnarray}
 \Delta \mathcal{W}(\textbf{x}) = \alpha^{2}\mathcal{W}(\textbf{x})\\
 \mathcal{U}''(u)+\frac{f'(u)}{f(u)}\mathcal{U}'(u)+\frac{\alpha^{2}}{u^{2}f(u)}\mathcal{U}(u)=0.\label{Eq56}
 \end{eqnarray}

From (\ref{Eq56}) we note that the radial function $ \mathcal{U}(u)$ near the boundary ($ u\rightarrow 0 $) of the $ AdS_{4} $ eventually becomes a constant,
\begin{eqnarray}
\mathcal{U}(u)|_{u\rightarrow 0}&=& \tilde{\mathcal{C}}= const.
\end{eqnarray}
On the other hand, the two dimensional function $ \mathcal{W}(\textbf{x}) $ takes the following form,
\begin{eqnarray}
\mathcal{W}(\textbf{x})&=&\varsigma(x,y) e^{-\sqrt{\alpha^{2}-\beta^{2}}x}e^{-\beta y}+\varsigma(-x,-y)e^{\sqrt{\alpha^{2}-\beta^{2}}x}e^{\beta y}
\end{eqnarray}
where,
\begin{eqnarray}
\varsigma(x,y) &=& 1 ~~~  for ~~~ x,y \geq 0\nonumber\\
&=& 0 ~~~ for~~~  x,y < 0.
\end{eqnarray}

Let us now consider the equations corresponding to the leading order in the CS coupling ($ \kappa $) which turn out to be,  
 \begin{eqnarray}
\mathcal{D}_t \mathcal{A}^{(1)(\kappa^{(1)})}_{t}+\frac{8 u^{2}f(u)}{r_+}\left[\tau'^{(\kappa^{(0)})}(u)\mathcal{F}^{(1)(\kappa^{(0)})}_{xy}(u,\textbf{x})+\mathcal{H}_{c2}\partial_u \theta^{(1)(\kappa^{(0)})}(u,\textbf{x}) \right] &=&\frac{2r_+^{2}}{u^{2}}\mathcal{Y}_{(t)}(u)\sigma (\textbf{x})\nonumber\\
\mathcal{D}_{s}\mathcal{A}^{(1)(\kappa^{(1)})}_{x}-8 r_+\left[\tau'^{(\kappa^{(0)})}(u)\partial_y \mathcal{A}^{(1)(\kappa^{(0)})}_{t}(u,\textbf{x})-\varphi'^{(\kappa^{(0)})}(u)\partial_y \theta^{(1)(\kappa^{(0)})}(u,\textbf{x}) \right]&=&\frac{r_+^{2}}{u^{2}}\mathcal{Y}_{(x)}(u)\epsilon_{x}^{y}\partial_y\sigma (\textbf{x})\nonumber\\
\mathcal{D}_{s}\mathcal{A}^{(1)(\kappa^{(1)})}_{y}-8 r_+\left[\varphi'^{(\kappa^{(0)})}(u)\partial_x \theta^{(1)(\kappa^{(0)})}(u,\textbf{x})-\tau'^{(\kappa^{(0)})}(u)\partial_x \mathcal{A}^{(1)(\kappa^{(0)})}_{t}(u,\textbf{x})\right]&=&\frac{r_+^{2}}{u^{2}}\mathcal{Y}_{(y)}(u)\epsilon_{y}^{x}\partial_x\sigma (\textbf{x})\nonumber\\ 
\mathcal{D}_{\theta}\theta^{(1)(\kappa^{(1)})}+\frac{4 u^{2}}{r_+}\left[ \mathcal{H}_{c2}\partial_u \mathcal{A}^{(1)(\kappa^{(0)})}_{t}(u,\textbf{x})+\varphi'^{(\kappa^{(0)})}(u)\mathcal{F}^{(1)(\kappa^{(0)})}_{xy}(u,\textbf{x})\right]&=&0\label{Eq60} 
\end{eqnarray}
where the functions $ \mathcal{Y}_{(t)}(u) $ and $ \mathcal{Y}_{(i)}(u) $ appearing on the r.h.s. of (\ref{Eq60}) have the following explicit forms,
\begin{eqnarray}
\mathcal{Y}_{(t)}(u)&=& 2 \varphi^{(\kappa^{(0)})}\Re^{(\kappa^{(0)})}\Re^{(\kappa^{(1)})}+\varphi^{(\kappa^{(1)})}\Re^{(\kappa^{(0)})}\nonumber\\
\mathcal{Y}_{(x)}(u)&=& \mathcal{Y}_{(y)}(u)= 2 \Re^{(\kappa^{(0)})}\Re^{(\kappa^{(1)})}. 
\end{eqnarray}

One should take a note on the fact that the functions $ \Re^{(\kappa^{(m)})} (u) $ ($ m\geq 0 $) solely depends on $ \varphi (u) $ (see (\ref{Eq40})) and hence could be treated as the source term for the fluctuations $ \mathcal{A}^{(1)(\kappa^{(1)})}_{t} $ and $ \mathcal{A}^{(1)(\kappa^{(1)})}_{i} $. This finally enables us to write down the solutions in the following form, 
\begin{eqnarray}
\mathcal{A}^{(1)(\kappa^{(1)})}_{t}&=& -  \int_{0}^{1}du'\int d\textbf{x}'\mathcal{K}(u',\textbf{x}') \mathcal{G}_{t}(u,u';\textbf{x},\textbf{x}')\nonumber\\
\mathcal{A}^{(1)(\kappa^{(1)})}_{i}&=& -r_+^{2}\int_{0}^{1}du'\int d\textbf{x}'\mathcal{Q}_{i}(u',\textbf{x}') \mathcal{G}_{s}(u,u';\textbf{x},\textbf{x}')\nonumber\\
\theta^{(1)(\kappa^{(1)})}&=&- \frac{4}{r_+}\int_{0}^{1}u'^{2}du'\int d\textbf{x}'\mathcal{N}(u',\textbf{x}') \mathcal{G}_{\theta}(u,u';\textbf{x},\textbf{x}')\label{Eq62}
\end{eqnarray} 
 where the various functions appearing in the integral expressions above in (\ref{Eq62}) have the following explicit functional forms,
 \begin{eqnarray}
 \mathcal{K}(u,\textbf{x})&=&\frac{2r_+^{2}}{u^{2}}\mathcal{Y}_{(t)}(u)\sigma (\textbf{x})-\frac{8 u^{2}f(u)}{r_+}\left[\tau'^{(\kappa^{(0)})}(u)\mathcal{F}^{(1)(\kappa^{(0)})}_{xy}(u,\textbf{x})+\mathcal{H}_{c2}\partial_{u} \theta^{(1)(\kappa^{(0)})}(u,\textbf{x}) \right]\nonumber\\
 \mathcal{N}(u,\textbf{x})&=& - \left[ \mathcal{H}_{c2}\partial_u \mathcal{A}^{(1)(\kappa^{(0)})}_{t}(u,\textbf{x})+\varphi'^{(\kappa^{(0)})}(u)\mathcal{F}^{(1)(\kappa^{(0)})}_{xy}(u,\textbf{x})\right]\nonumber\\
 \mathcal{Q}_{i}(u,\textbf{x})&=&\epsilon_{i}^{j}\partial_j \Xi (u,\textbf{x})\nonumber\\
  \Xi (u,\textbf{x})&=&\frac{\mathcal{Y}_{(i)}(u)}{u^{2}}\sigma(\textbf{x})+\frac{8}{r_+}\left[\tau'^{(\kappa^{(0)})}(u) \mathcal{A}^{(1)(\kappa^{(0)})}_{t}(u,\textbf{x})-\varphi'^{(\kappa^{(0)})}(u) \theta^{(1)(\kappa^{(0)})}(u,\textbf{x}) \right].
 \end{eqnarray}
 
Once the explicit solutions for the fluctuations (\ref{Eq51}) and (\ref{Eq62}) are in hand, one can then further proceed to calculate the $ U(1) $ current for the boundary theory corresponding to the leading order fluctuations in the scalar field   profile, namely $ \Psi =\psi_{1} $ which is the goal of the next section.

\subsection{Computation of $ J_i $ }

In the language of the AdS/CFT duality, the spontaneous breaking of the \textit{local} $ U(1) $ symmetry in the bulk due to the formation of the non trivial scalar hair ($ \Psi $) eventually breaks the \textit{global} $ U(1) $ symmetry for the boundary theory. In this sense the symmetry broken phase of the boundary theory corresponds to a super-fluid phase rather than a superconducting phase. In our computations of this section we shall show that the global $ U(1) $ current computed from the standard holographic prescription can be expressed as a local function of the vortex solution of the type (\ref{Eq24}) that we have found earlier in the context of Maxwell-CS type vortices. This finally indicates that there exists a dual gravitational description for the abelian CS Higgs model in ($ 2+1 $) dimensions.

Following AdS/CFT prescription \cite{Hartnoll:2009sz}, the $ U(1) $ current could be computed as,
\begin{eqnarray}
\langle J^{\mu}\rangle = \frac{\delta S_{(os)}}{\delta A_{\mu}}\mid_{u\rightarrow 0}=\sqrt{-g}F^{\mu u}+4\kappa \theta \epsilon^{\mu u \rho\sigma}F_{\rho\sigma}\mid_{u\rightarrow 0}\label{eqJ}.
\end{eqnarray}

The spatial component of the current that immediately follows from (\ref{eqJ}) has the following form,
\begin{eqnarray}
\langle J_{i}\rangle =  \mathcal{F}_{iu}^{(1)(\kappa^{(0)})}+\kappa \left[\mathcal{F}_{iu}^{(1)(\kappa^{(1)})}+4 \epsilon^{iu\rho\sigma}\left(\theta^{(0)(\kappa^{(0)})}\mathcal{F}_{\rho\sigma}^{(1)(\kappa^{(0)})} +\theta^{(1)(\kappa^{(0)})}\mathcal{F}_{\rho\sigma}^{(0)(\kappa^{(0)})}\right)\right]+\mathcal{O}(\varepsilon).\label{Eq65}
\end{eqnarray}
Here we have scaled both sides by a factor of $ \varepsilon $. Finally, substituting (\ref{Eq51}) and (\ref{Eq62}) in to (\ref{Eq65}) we get the following expression for the current namely,
\begin{eqnarray}
\langle J_{i}\rangle = \epsilon_{i}^{j}\partial_j \Omega (\textbf{x})\mid_{u\rightarrow 0}\label{Eq66}
\end{eqnarray}
where\footnote{We have absorbed the overall $ r_+^{2} $ factor.} the function $ \Omega (\textbf{x}) $ could be expressed as,
\begin{eqnarray}
\Omega (\textbf{x})=\int_{0}^{1} \frac{du'}{u'^{2}}\Re^{2(\kappa^{(0)})}(u')\partial_u\int d\textbf{x}' \mathcal{G}_{s}(u,u';\textbf{x},\textbf{x}') \sigma (\textbf{x}')\mid_{u\rightarrow 0}+\kappa \Pi (u,x)\mid_{u\rightarrow 0}.\label{Eq67}
\end{eqnarray}
The functional form of $ \Pi (u,x) $ is given by the following expression,
\begin{eqnarray}
\Pi (u,x) =\int_{0}^{1}du'\partial_u \int d\textbf{x}'\Xi(u',\textbf{x}')\mathcal{G}_{s}(u,u';\textbf{x},\textbf{x}')\nonumber\\
-16  \mathcal{C}_{1}\int_{0}^{1}du' \frac{\varphi^{(\kappa^{(0)})}(u')}{u'^{2}}\Re^{2(\kappa^{(0)})}(u')\int d\textbf{x}' \mathcal{G}_{t}(u,u';\textbf{x},\textbf{x}')\sigma (\textbf{x}').
\end{eqnarray}

From (\ref{Eq67}) it is quite evident that $ \Omega (\textbf{x}) $ is a \textit{non local} function of the spatial coordinates ($ \textbf{x} $). This is because of the fact that in order to compute $ \Omega (\textbf{x}) $ one needs to perform an integration over an infinitesimal region surrounding the point $ \textbf{x} $. On the other hand the GL form of the supercurrent that we have obtained for the mixed Maxwell-CS type vortices is a \textit{local} function of the vortex solutions (see (\ref{Eq24}). Therefore somehow one has to bring the above non local form (\ref{Eq66}) of the supercurrent in to a local form. This is usually achieved by considering the so called \textit{long wave length} limit where one performs the integration following certain suitable approximations \cite{Maeda:2009vf}. In the following we describe the entire procedure set by step. As a first step, we split the Green's function in to two pieces as follows \cite{Maeda:2009vf},
\begin{eqnarray}
\mathcal{G}_{t}(u,u';\textbf{x},\textbf{x}')&=&\sum_{\upsilon}\alpha_{\upsilon}(u)\alpha_{\upsilon}^{\dagger}(u')\tilde{\mathcal{G}_{t}}(\textbf{x}-\textbf{x}',\upsilon)\nonumber\\
\mathcal{G}_{s}(u,u';\textbf{x},\textbf{x}')&=&\sum_{\Lambda}\beta_{\Lambda}(u)\beta_{\Lambda}^{\dagger}(u')\tilde{\mathcal{G}_{s}}(\textbf{x}-\textbf{x}',\Lambda)
\end{eqnarray}
where $ \alpha_{\upsilon}(u) $ and $ \beta_{\Lambda}(u) $ are some radial functions that satisfy the following eigen value equations namely,
\begin{eqnarray}
  D_{t}\alpha_{\upsilon}(u)=\upsilon
 \alpha_{\upsilon}(u);~~~\sum\limits_{\upsilon}\alpha_{a}(u)\alpha_{\upsilon}^{\dagger}(u')=\delta(u-u');~~~\langle\alpha_{\upsilon}|\alpha_{\upsilon'}\rangle 
  = \delta_{\upsilon \upsilon'}\nonumber\\ 
 D_{s}\beta_{\Lambda}(u)=\Lambda \beta_{\Lambda}(u);~~~\sum\limits_{\Lambda}\beta_{\Lambda}(u)\beta_{\Lambda}^{\dagger}(u')=\delta(u-u');~~~\langle\beta_{\Lambda}|\beta_{\Lambda'}\rangle = \delta_{\Lambda \Lambda'}.
\end{eqnarray}
where, $  D_{t} = - u^{2}f(u)\partial_{u}^{2} $ and $  D_{s} = - \partial_u (u^{2}f(u)\partial_{u}) $ are the differential operators that solely depend on the radial coordinates.

$\tilde{\mathcal{G}}_{t}(\textbf{x}-\textbf{x}',\upsilon) $ and $\tilde{\mathcal{G}}_{s}(\textbf{x}-\textbf{x}',\Lambda)$ are the Green's functions defined over the two dimensional spatial hyper surface that satisfy the equation of the following type \cite{Maeda:2009vf},
\begin{eqnarray}
(\Delta - \varrho^{2})\tilde{\mathcal{G}}(\textbf{x},\varrho^{2}) &=& -\delta (\textbf{x}).\label{eqn25}
\end{eqnarray}
 The solution for (\ref{eqn25}) could be expressed in terms of modified Bessel's function that takes the following form\footnote{It is customary to mention that the two dimensional Green's function satisfies the limiting condition namely $ lim_{|\textbf{x}|\rightarrow \infty}|\tilde{\mathcal{G}}(\textbf{x})|<\infty $.},
 \begin{eqnarray}
\tilde{\mathcal{G}}(\textbf{x},\varrho^{2})=\frac{1}{2\pi}K_0 (\varrho |x|).
\end{eqnarray}

Since the modified Bessel's function has a sharp fall off whose width is roughly given by $ \frac{1}{\varrho} $, therefore the spatial Green's functions namely $\tilde{\mathcal{G}}_{t}(\textbf{x}-\textbf{x}',\upsilon) $ (or, $\tilde{\mathcal{G}}_{s}(\textbf{x}-\textbf{x}',\Lambda)$) are also associated with a natural length scale of the order $ \frac{1}{\sqrt{\upsilon}} $ (or, $ \frac{1}{\sqrt{\Lambda}} $). For $ \mid x\mid \gg \frac{1}{\sqrt{\upsilon}} $ (or $ \frac{1}{\sqrt{\Lambda}} $) these spatial Green's functions die out rapidly. Therefore the theory we are concerned with has an additional length scale ($ \frac{1}{\sqrt{\upsilon}} $ or $ \frac{1}{\sqrt{\Lambda}} $) apart from the coherence length ($ \xi $) mentioned earlier (see the discussion below (\ref{Eq45})).

\underline{\emph{Long wavelength limit}}: By long wavelength limit we mean $ \frac{1}{\sqrt{\upsilon}} $ or ($\frac{1}{\sqrt{\Lambda}}$) $ \ll \xi $. This implies that the condensate ($ \sigma(\textbf{x}) $) remains almost uniform over the scale in which the spatial Green's functions fluctuates, i.e; $ \mid \frac{\Delta \sigma} {\Delta \textbf{x}}\mid_{\textbf{x}=\frac{1}{\sqrt{\upsilon}},\frac{1}{\sqrt{\Lambda}}}  \ll 1 $ . 
This further implies that one can in fact expand the vortex solution in a Taylor series about $ \textbf{x}=\textbf{x}' $ as\footnote{Here $ \mid \textbf{x}-\textbf{x}'\mid\approx \frac{1}{\sqrt{\upsilon}} $ (or $ \frac{1}{\sqrt{\Lambda}} $).},
\begin{eqnarray}
\sigma (\textbf{x}')&=&\sigma (\textbf{x}) +\mathcal{O}(\frac{1}{\sqrt{\upsilon}})\nonumber\\
\sigma (\textbf{x}')&=&\sigma (\textbf{x}) +\mathcal{O}(\frac{1}{\sqrt{\Lambda}}).\label{Eq74}
\end{eqnarray} 

Using (\ref{eqn25}) and (\ref{Eq74}) it is quite straightforward to show that, 
\begin{eqnarray}
\int d\textbf{x}'\tilde{\mathcal{G}_{t}}(\textbf{x}-\textbf{x}',\upsilon)\sigma(\textbf{x}')&=&\frac{\sigma(\textbf{x})}{\upsilon}+\mathcal{O}(\frac{1}{\upsilon^{3/2}})\nonumber\\
\int d\textbf{x}'\tilde{\mathcal{G}_{s}}(\textbf{x}-\textbf{x}',\Lambda)\sigma(\textbf{x}')&=&\frac{\sigma(\textbf{x})}{\Lambda}+\mathcal{O}(\frac{1}{\Lambda^{3/2}}).\label{Eq75}
\end{eqnarray}

Substituting (\ref{Eq75}) in to (\ref{Eq67}) we finally arrive at the following local expression for the supercurrent namely,
\begin{eqnarray}
\langle J_{i}\rangle \sim \mathcal{P}\epsilon_{i}^{j}\partial_j\sigma (\textbf{x})+\kappa \left( \sum_{i=1}^{2}\mathcal{P}_{i}\right)\epsilon_{i}^{j}\partial_j\sigma (\textbf{x})\label{Eq76}
\end{eqnarray}
where the coefficients $ \mathcal{P} $, $ \mathcal{P}_{1} $ and $ \mathcal{P}_{2} $ are some numerical factors that have the following explicit forms,
\begin{eqnarray}
\mathcal{P}&=& \sum_{\Lambda}\frac{\beta_{\Lambda}'(0)}{\Lambda}\int_{0}^{1} \frac{du'}{u'^{2}}\Re^{2(\kappa^{(0)})}(u')\beta_{\Lambda}^{\dagger}(u')\nonumber\\
\mathcal{P}_{1}&=&\sum_{\Lambda}\frac{\beta_{\Lambda}'(0)}{\Lambda}\int_{0}^{1} \frac{du'}{u'^{2}}\mathcal{Y}_{(i)}(u')\beta_{\Lambda}^{\dagger}(u')\nonumber\\
\mathcal{P}_{2}&=&-16r_{+}\sum_{\upsilon,\Lambda}\frac{\beta_{\Lambda}'(0)}{\Lambda\upsilon}\int_{0}^{1}du'\alpha_{\upsilon}(u')\tau'^{(\kappa^{(0)})}(u')\beta^{\dagger}_{\Lambda}(u')\int_{0}^{1}du''\frac{\varphi^{(\kappa^{(0)})}(u'')}{u''^{2}}\Re^{2(\kappa^{(0)})}(u'')\alpha^{\dagger}_{\upsilon}(u'').\nonumber\\
\end{eqnarray}

From (\ref{Eq76}) it is indeed quite evident that the \textit{local} form of the global $ U(1) $ current computed at the boundary of the $ AdS_{4} $ is remarkably identical to that with the local expression for the suepercurrent associated with the Maxwell-CS type vortices that has been found earlier in (\ref{Eq24}). In both the cases the \textit{local} structure of the supercurrent is of the following form,
\begin{eqnarray}
j_{MCS}\sim j_{GL} + \kappa (\#) j_{GL}.
\end{eqnarray}
Based on this structural similarity of the supercurrent in both the cases, we therefore conclude that the abelian Higgs model coupled to $ AdS_{4} $ in the presence of a coupling of the form $ \theta F\wedge F $ could be regarded as the dual gravitational description for the Maxwell-CS type vortices in ($ 2+1 $) dimensions. Our claim is compatible with the earlier findings in \cite{ref36}. The basic difference between the present analysis to that with the earlier observation is that in the present analysis we have established the holographic connection based on the structural similarity in the local form of the GL current associated with the Maxwell-CS type vortices which was missing in the earlier literature. In the subsequent sections we shall exploit this duality a bit further in order to gain some more physical insights regarding Maxwell-CS type vortices in ($ 2+1 $) dimensions.

\section{Free energy}

Computation of free energy is important in order to determine the thermodynamic stability of any given configuration. In this section, based on the gauge/gravity duality principle, we compute the free energy for the Maxwell-CS vortex configuration in ($ 2+1 $) dimensions. It has been observed that the configuration with vortices possesses lower free energy which therefore corresponds to a thermodynamically stable configuration. 

In the holographic framework, the free energy for the boundary theory could be formally expressed in terms of the \textit{on-shell} gravity action defined in the bulk,
\begin{eqnarray}
F=-S_{(os)}.
\end{eqnarray}

Considering a static field configuration and taking in to account the boundary behaviour for the scalar fields ($ \Psi \sim c u^{2} $), one can easily show that for the present case of study, the \textit{on-shell} action corresponding to the scalar field vanishes identically,
\begin{equation}
S_{\psi}|_{(os)}=-\frac{1}{2}\int_{\partial \Sigma}d\Sigma_{\mu}\sqrt{-g}(\nabla^{\mu}-iA^{\mu})|\Psi|^{2}=0.
\end{equation}

Let us now evaluate the remaining part of the action that consists of $ U(1) $ gauge field and its coupling to the neutral pseudo scalar field ($ \theta $). In order to do that, as a first step, we expand the action perturbatively in the parameter ($ \varepsilon_{H} $) namely,
\begin{eqnarray}
S_{(os)}= S_{(os)}^{(0)}+\varepsilon_{H} S_{(os)}^{(1)}+\varepsilon_{H}^{2}S_{(os)}^{(2)}+\mathcal{O}(\varepsilon_{H}^{3}).\label{Eq81}
\end{eqnarray}

The first term on the r.h.s of the above expansion (\ref{Eq81}) corresponds to a trivial configuration i.e; the configuration without any scalar condensate ($ \Psi=0 $). On the other hand, all the subsequent terms in the expansion (\ref{Eq81}) contains the information corresponding to some non zero condensate ($ \Psi\neq 0 $). Let us first consider the on-shell action corresponding to the leading order in fluctuations. Using the equations of motion (\ref{Eq33}) one finds,
\begin{eqnarray}
S^{(1)}_{(os)}= -\int_{\partial \Sigma} d\Sigma_{u}\sqrt{-g}\left[F^{(0)u\mu}A^{(1)}_{\mu}+\frac{4\kappa}{\sqrt{-g}}\epsilon^{u\mu\rho\sigma}F^{(0)}_{\rho\sigma}A^{(1)}_{\mu} \right]_{u=0}\nonumber\\
-2\int_{\partial \Sigma} d\Sigma_{u}\sqrt{-g}\nabla^{u}\theta^{(0)}\theta^{(1)}\mid_{u=0}=0.\label{Eq82}
\end{eqnarray}
The first term on the r.h.s of (\ref{Eq82}) vanishes because we have considered our boundary theory at some fixed value of the chemical potential ($ \mu $) whereas on the other hand the second term vanishes because of the fact that $ \theta^{(0)}\rightarrow \mathcal{C}_{1} $ near the boundary of the $ AdS_{4} $ (see (\ref{theta})).

Therefore we may hope that the first non trivial correction to the free energy might appear at the quadratic order in $ \varepsilon_{H} $. After using equations of motion (\ref{Eq33}) and (\ref{Eq48}) and considering the orthogonality relation\footnote{By orthogonality condition we mean $\int_{\Sigma}d^{4}x \sqrt{-g}A_{\mu}^{(1)}j^{\mu(1)}=0$.} the action at the quadratic level in fluctuations turns out to be,
\begin{eqnarray}
S^{(2)}_{(os)}= -\int_{\partial \Sigma} d\Sigma_{u}\frac{\sqrt{-g}}{2}\left[F^{(1)u\mu}A^{(1)}_{\mu}+\frac{4\kappa}{\sqrt{-g}}\epsilon^{u\mu\rho\sigma}\left(\theta^{(0)}F^{(1)}_{\rho\sigma}+\theta^{(1)}F^{(0)}_{\rho\sigma} \right)A^{(1)}_{\mu}  \right]_{u=0}\nonumber\\
-\int_{\partial \Sigma} d\Sigma_{u}\sqrt{-g}\left[F^{(0)u\mu}A^{(2)}_{\mu}+\frac{4\kappa}{\sqrt{-g}}\epsilon^{u\mu\rho\sigma}F^{(0)}_{\rho\sigma}A^{(2)}_{\mu} \right]_{u=0}
 -\int_{\partial \Sigma} d\Sigma_{u}\sqrt{-g}\left[ 2\nabla^{u}\theta^{(0)}\theta^{(2)}+\nabla^{u}\theta^{(1)}\theta^{(1)}\right]_{u=0}.\label{Eq83}\nonumber\\
\end{eqnarray}
Following the previous arguments the last two terms on the r.h.s of (\ref{Eq83}) vanishes identically. On the other hand using (\ref{eqJ}) the first two terms could be combined together to get some finite answer for the free energy of the Maxwell-CS type vortices in ($ 2+1 $) dimensions namely, 
\begin{eqnarray}
F=-\frac{\varepsilon^{2}}{2}\int_{\Re^{2}} d\textbf{x} \langle J^{i}\rangle a_{i}(\textbf{x})=-\frac{\varepsilon^{2}\mathcal{H}_{c2}}{2}\int_{\Re^{2}} d\textbf{x} \Omega (\textbf{x}).
\end{eqnarray}

If we consider $ V $ to be the volume of some bounded region over the two dimensional hyper surface ($ \Re^{2} $), then the free energy density turns out to be,
\begin{eqnarray}
F/V= -\frac{\varepsilon^{2}\mathcal{H}_{c2}}{2} \langle\Omega (\textbf{x}) \rangle
\end{eqnarray}
where $ \langle\Omega (\textbf{x}) \rangle $ stands for the average of $ \Omega (\textbf{x}) $ over the two dimensional hyper surface ($ \Re^{2} $).
Since $ F/V $ turns out to be negative therefore we may cnclude that the vortex configuration is stable over the trivial configuration where there is no charge condensate ($ \Psi =0 $).

\section{More about the duality}

Up till now we have seen that starting with the abelian Higgs model coupled to gravity in $ AdS_{4} $ space time in the presence of a $ \theta F\wedge F $ type coupling one can in fact reproduce the GL current associated with Maxwell-CS type vortices that is originally found in ($ 2+1 $) dimensions. More over based on the holographic calculations one can further show that the configuration with CS vortices is a stable one. Motivated from these observations, in this section we shall exploit this duality further in order to calculate the coherence length ($ \xi $) as well as the magnetic penetration depth ($ \delta $) for the Maxwell-CS vortex configuration. 


\subsection{Coherence length}
Coherence length ($ \xi $) is defined as the pole of the two point correlation of the order parameter in the momentum space \cite{Maeda:2008ir}-\cite{Zeng:2009dr},
\begin{eqnarray}
 \langle \mathcal{O}(k)\mathcal{O}(-k)\rangle\sim \frac{1}{|k|^{2}+\frac{1}{\xi^{2}}}.
\end{eqnarray}
 
 Solutions to the eigen value problem associated with the quasi normal modes corresponding to some static perturbations in the bulk space time essentially gives the pole of the static correlation function for the boundary theory. In order to get the pole one needs to satisfy the condition $ - k^{2}=\frac{1}{\xi^{2}} $. Since near the critical point ($ T=T_c $) the coherence length ($ \xi $) diverges, therefore our aim would be to solve the eigenvalue problem with the condition\footnote{Here $ \varepsilon_{T}=\frac{T_c-T}{T_c} $, with the condition that $ \mid \varepsilon_{T}\mid\ll 1 $.} $ \lim_{\varepsilon_{T}\rightarrow 0}-k^{2}=-k_{\ast}^{2}=\frac{1}{\xi^{2}}=0 $.

In order to start our analysis we take the following ansatz for the gauge field ($ A_{\mu} $), scalar field ($ \Psi $) and the neutral pseudo scalar field ($ \theta $) namely,
\begin{eqnarray}
A_{\mu}=(\Phi(u),0,0,\mathcal{H}_{c2}x),~~~\Psi = \Psi (u),~~~\theta = \theta (u).\label{Eq87}
\end{eqnarray} 
 
 Substituting the ansatz (\ref{Eq87}) in to (\ref{Eq30}) we arrive at the following set of equations,
\begin{eqnarray}
\partial_u^{2} \Phi +\frac{8\kappa \mathcal{H}_{c2}}{r_+}\partial_u \theta -\frac{2r_+^{2}}{u^{4}f(u)}\Psi^{2}\Phi &=& 0\nonumber\\
\left[ \partial_u \left(f(u)\partial_u \right)-\frac{m^{2}r_+^{2}}{u^{4}}\right]\Psi +\frac{r_+^{2}\Phi^{2}}{u^{4}f(u)}\Psi -\frac{\mathcal{H}_{c2}^{2}x^{2}}{u^{2}}\Psi &=& 0\nonumber\\
\partial_u \left(f(u)\partial_u \right)\theta + \frac{4\kappa \mathcal{H}_{c2}}{r_+}\partial_u \Phi &=& 0\label{eq41}.  
\end{eqnarray}
 
 Following our previous approach, we would like to solve the above set of equations (\ref{eq41}) perturbatively in the fluctuations of the scalar field namely,
\begin{eqnarray}
 \Phi &=& \varphi(u)+\varepsilon_{T} A^{(1)}_{t}(u)+\mathcal{O}(\varepsilon_{T}^{2})\nonumber\\
 \theta &=& \tau(u)+\varepsilon_{T} \theta^{(1)}(u)+\mathcal{O}(\varepsilon_{T}^{2})\nonumber\\
 \Psi &=& \varepsilon_{T}^{1/2}\psi_{1}(u)+\varepsilon_{T}^{3/2}\psi_{2}(u)+\mathcal{O}(\varepsilon_{T}^{5/2})\label{eq43}.
 \end{eqnarray}

Our next aim would be to write down the equations for the static fluctuations in the leading order which could be enumerated as,
\begin{eqnarray}
\partial_u^{2} A^{(1)}_{t}+\frac{8\kappa \mathcal{H}_{c2}}{r_+}\partial_u \theta^{(1)}-\frac{2r_+^{2}}{u^{4}f(u)}\psi_1^{2}\varphi &=& 0\nonumber\\
\left[ \partial_u \left(f(u)\partial_u \right)-\frac{m^{2}r_+^{2}}{u^{4}}\right]\psi_{1} +\frac{r_+^{2}\varphi^{2}}{u^{4}f(u)}\psi_1 -\frac{\mathcal{H}_{c2}^{2}x^{2}}{u^{2}}\psi_1 &=& 0\nonumber\\
\partial_u \left(f(u)\partial_u \right)\theta^{(1)} + \frac{4\kappa \mathcal{H}_{c2}}{r_+}\partial_u A^{(1)}_{t} &=& 0\label{eq44}.
\end{eqnarray}

Let us now consider some liner perturbation of the gauge field, scalar field and the neutral pseudo scalar field in the presence of a non zero spatial momentum along the $ x $ direction which reads as,
\begin{eqnarray}
\delta A_{\mu}&=& [a_{t}(u,k)dt+a_{x}(u,k)dx+a_{y}(u,k)dy]e^{ik.x}\nonumber\\
\delta\Psi &=& [b(u,k)+i \hat{b}(u,k)]e^{ik.x}\nonumber\\
\delta\theta &=&  c(u,k)e^{ik.x}\label{Eq91}
\end{eqnarray}
where $ |a|, |b|, |c|\ll 1 $.

At this point it is customary to mention that for $ m^{2}= -2 $, the asymptotic behavior of $ a_{t}(u,k) $, $ b(u,k) $ and $ c(u,k) $ could be enumerated as,
\begin{eqnarray}
a_{t}(u,k)|_{u=1}&=& 0\nonumber\\
b(u,k)|_{u=1}&=& regular\nonumber\\
a_{t}(u,k)|_{u\rightarrow 0} &=& (const.)\times u\nonumber\\
b(u,k)|_{u\rightarrow 0} &=& (const.)\times u^{2}\nonumber\\
c(u,k)|_{u\rightarrow 0} &=& const. \label{eq42}
\end{eqnarray}

From (\ref{Eq30}) it is now quite straight forward to obtain the equations of motion corresponding to the above static perturbations (\ref{Eq91}) which could be enumerated as follows,
\begin{eqnarray}
k^{2}a_t &=& \left( u^{2}f(u)\partial_u^{2}- \frac{2 r_+^{2}\Psi^{2}}{u^{2}}\right) a_t (u,k) +\frac{8\kappa u^{2} f(u)}{r_+}\left(\mathcal{H}_{c2}\partial_u c +i k a_y \partial_u\theta  \right) -\frac{4b r_+^{2}}{u^{2}}\Psi\Phi\nonumber\\
k^{2}b &=& \left(u^{2}\partial_u \left(f(u)\partial_u \right) +\frac{r_+^{2}\Phi^{2}}{u^{2}f(u)} - \frac{m^{2}r_+^{2}}{u^{2}}-\mathcal{H}_{c2}^{2}x^{2} \right)b(u,k) +\frac{2 a_t r_+^{2}}{u^{2}f(u)} \Phi\Psi -2\mathcal{H}_{c2}x a_y \Psi\nonumber\\
0&=& \left[\partial_u \left(u^{2}f(u)\partial_u \right)+\frac{2r_+^{2}}{u^{2}}\Psi^{2}\right]a_x(u,k)\nonumber\\ 
k^{2}a_y &=& \left[ \partial_u \left(u^{2}f(u)\partial_u \right)-\frac{2r_+^{2}\Psi^{2}}{u^{2}}\right]a_y(u,k) +i8k\kappa r_+ \left( a_t \partial_u \theta -c \partial_u \Phi \right) - \frac{4 b r_+^{2}\mathcal{H}_{c2}x}{u^{2}}\Psi\nonumber\\
k^{2}c &=& u^{2}\partial_u \left(f(u)\partial_u \right)c(u,k)+\frac{4\kappa u^{2}}{r_+}\left(\mathcal{H}_{c2}\partial_u a_t + i k a_{y}\partial_u \Phi \right).\label{eq46} 
\end{eqnarray}

Let us now focus on the first two equations of (\ref{eq46}). In order to proceed further we first Wick rotate the wave vector $ k $ namely $k \rightarrow i k $. At the end of the calculation one has to get back to the original wave vector $ k $ through an inverse Wick rotation. With this operation of Wick rotating the wave vector, the first two equations of (\ref{eq46}) turn out to be,  
\begin{eqnarray}
-\tilde{k}^{2}a_t &=& \left( u^{2}f(u)\partial_u^{2}- \frac{2 r_+^{2}\Psi^{2}}{u^{2}}\right) a_t (u,\tilde{k})+\frac{8\kappa u^{2} f(u)}{r_+}\mathcal{H}_{c2}\partial_u c -\frac{4b r_+^{2}}{u^{2}}\Psi\Phi\nonumber\\
-\tilde{k}^{2} b &=&\left(u^{2}\partial_u \left(f(u)\partial_u \right) +\frac{r_+^{2}\Phi^{2}}{u^{2}f(u)} - \frac{m^{2}r_+^{2}}{u^{2}}-\mathcal{H}_{c2}^{2}x^{2} \right)b(u,\tilde{k}) +\frac{2 a_t r_+^{2}}{u^{2}f(u)} \Phi\Psi -2\mathcal{H}_{c2}x a_y \Psi\label{eq47}\nonumber\\
\end{eqnarray}
where we have defined,
\begin{eqnarray}
\tilde{k}= k- \frac{4\kappa u^{2}f(u)}{r_+}\frac{a_y}{a_t}\partial_u \theta. \label{eq48}
\end{eqnarray}
In order to arrive at the above set of equations (\ref{eq47}) we have assumed that the ratio $\mid \frac{a_y}{a_t}\mid \ll 1 $, which means that for our case of study the ratio of the perturbations for the $ y $ component of the gauge field to that with the temporal component should be very small so that we can always ignore the term $\sim \mathcal{O} (\kappa b\mid \frac{a_y}{a_t}\mid)\ll 1 $ compared to unity. Since we have already neglected all the perturbations at the quadratic level and considered only the linear perturbations in (\ref{Eq91}), therefore this seems to be a valid approximation.  Furthermore at this point one should also take a note on the fact that since $ k^{2} $ itself vanishes at the critical point ($ T=T_c $), therefore from the above relation (\ref{eq48}) it is easy to note that $ \lim_{\varepsilon_{T}\rightarrow 0}\tilde{k}^{2}=\lim_{\varepsilon_{T}\rightarrow 0}k^{2}=\mathcal{O}(\kappa^{2}) =0$. This implies that by solving the eigen value equation for the newly defined wave vector ($ \tilde{k} $) one can in fact get the correct expression for the coherence length ($ \xi $) near the critical point of the phase transition line. 

Substituting (\ref{eq43}) in to (\ref{eq47}) we find,

\begin{eqnarray}
-\tilde{k}^{2}a_t &=& \left( u^{2}f(u)\partial_u^{2}- \frac{2 \varepsilon_{T} r_+^{2}\psi_1^{2}}{u^{2}}\right) a_t (u,\tilde{k})+\frac{8\kappa u^{2} f(u)}{r_+}\mathcal{H}_{c2}\partial_u c -\frac{4\sqrt{\varepsilon_{T}} r_+^{2}\psi_1 \varphi}{u^{2}} b(u,\tilde{k})\nonumber\\
-\tilde{k}^{2} b &=&\left(\mathcal{D}_{b} +\frac{2\varepsilon_{T} r_+^{2}\varphi A^{(1)}_{t}}{u^{2}f(u)} \right)b(u,\tilde{k}) +\frac{2\sqrt{\varepsilon_{T}} a_t r_+^{2}}{u^{2}f(u)} \varphi\psi_1 -2\sqrt{\varepsilon_{T}}\mathcal{H}_{c2}x a_y \psi_1 \label{eq49}
\end{eqnarray}
where,
\begin{eqnarray}
\mathcal{D}_{b}= u^{2}\partial_u \left(f(u)\partial_u \right) +\frac{r_+^{2}\varphi^{2}}{u^{2}f(u)} - \frac{m^{2}r_+^{2}}{u^{2}}-\mathcal{H}_{c2}^{2}x^{2}.
\end{eqnarray}

Let us first note down the zeroth order solutions to the above set of equations (\ref{eq49}) that are also consistent with (\ref{eq10}) and (\ref{eq42}),
\begin{eqnarray}
b^{(0)}=\psi_1 , ~~~, a_t^{(0)} = 0 , ~~~, c^{(0)}=0 , ~~~,a_y^{(0)}=0.
\end{eqnarray}

These solutions are perfectly consistent with the fact that at the zeroth order level the gauge field ($ A_{\mu} $) as well as the neutral pseudo scalar field ($ \theta $) are solely functions of radial coordinate ($ u $) (see (\ref{eq10})). On the other hand the perturbations defined in (\ref{Eq91}) contain some explicit fluctuations with spatial momentum along the $ x $ direction.
 Therefore from (\ref{eq49}) we note that the first non trivial correction appears that appears to these perturbations are in the $ \sqrt{\varepsilon_{T}} $ order. For our case we take,
\begin{eqnarray}
a_t = \sqrt{\varepsilon_{T}}\tilde{a}_t , ~~~c=\sqrt{\varepsilon_{T}} \tilde{c} , ~~~a_y = \sqrt{\varepsilon_{T}}\tilde{a}_{y}\label{eq52}.
\end{eqnarray}

Substituting (\ref{eq52}) in to (\ref{eq49}) we find,
\begin{eqnarray}
-\tilde{k}^{2}\tilde{a}_t &=& \left( u^{2}f(u)\partial_u^{2}\tilde{a}_t (u,\tilde{k})-\frac{4 r_+^{2}\psi_1 \varphi}{u^{2}} b(u,\tilde{k})+\frac{8\kappa u^{2} f(u)}{r_+}\mathcal{H}_{c2}\partial_u \tilde{c}\right)- \frac{2 \varepsilon_{T} r_+^{2}\psi_1^{2}}{u^{2}}\tilde{a}_t\nonumber\\
 -\tilde{k}^{2} b &=&\mathcal{D}_{b}b(u,\tilde{k}) +\varepsilon_{T}\left( \frac{2 b r_+^{2}\varphi A^{(1)}_{t}}{u^{2}f(u)}  +\frac{2 \tilde{a}_t r_+^{2}}{u^{2}f(u)} \varphi\psi_1 -2\mathcal{H}_{c2}x \tilde{a}_y \psi_1\right)\label{eq53}. 
\end{eqnarray}

Moreover substituting (\ref{eq52}) in to the last two equations of (\ref{eq46}) we note that,
\begin{eqnarray}
-k^{2}\tilde{a}_y &=&  \partial_u \left(u^{2}f(u)\partial_u \right)\tilde{a}_y(u,k) - 8k\kappa r_+ \left( \tilde{a}_t \partial_u \tau -\tilde{c} \partial_u \varphi \right) - \frac{4 r_+^{2}\psi_{1}^{2}\mathcal{H}_{c2}x}{u^{2}}+\mathcal{O}(\varepsilon_{T})\nonumber\\
-k^{2}\tilde{c} &=& u^{2}\partial_u \left(f(u)\partial_u \right)\tilde{c}(u,k)+\frac{4\kappa u^{2}}{r_+}\left(\mathcal{H}_{c2}\partial_u \tilde{a}_t - k \tilde{a}_{y}\partial_u \varphi \right)+\mathcal{O}(\varepsilon_{T}).\label{Eq101}
\end{eqnarray}

Looking at the above set of equations (\ref{eq53}) and (\ref{Eq101}) it is now quite intuitive to write down the following set of expansion for the coefficients of the static perturbations namely,
\begin{eqnarray}
b &=& \psi_1 + \varepsilon_{T} b^{(1)} + \mathcal{O}(\varepsilon_{T}^{2})\nonumber\\
\tilde{a}_t &=& \tilde{a}^{(0)}_t + \mathcal{O}(\varepsilon_{T})\nonumber\\
\tilde{a}_y &=& \tilde{a}^{(0)}_y + \mathcal{O}(\varepsilon_{T})\nonumber\\
\tilde{c}&=& \tilde{c}^{(0)}+ \mathcal{O}(\varepsilon_{T})\nonumber\\
 \tilde{k}_{\ast}^{2}&=& \varepsilon_{T} \tilde{k}_{1}^{2}+ \mathcal{O}(\varepsilon_{T}^{2})\label{eq55}.
\end{eqnarray}

In next step, we are going to take the $ \varepsilon_{T}\rightarrow 0 $
limit of (\ref{eq53}), which eventually means that $ \tilde{k}^{2}\rightarrow \tilde{k}_{\ast}^{2}$. Considering this limit and using (\ref{eq44}) we arrive at the following set of equations namely,
\begin{eqnarray}
\partial_u^{2}\tilde{a}^{(0)}_t (u,\tilde{k}_{\ast})&=& \frac{4r_+^{2}\psi_{1}^{2}\varphi}{u^{4}f(u)}-\frac{8\kappa}{r_+}\mathcal{H}_{c2}\partial_u \tilde{c}^{(0)}= 2\partial_u^{2} A^{(1)}_{t}+\frac{8\kappa \mathcal{H}_{c2}}{r_+}\left( 2\partial_u \theta^{(1)} -\partial_u \tilde{c}^{(0)}\right)\nonumber\\
-\tilde{k}_{1}^{2} \psi_{1} &=&\mathcal{D}_{b}b^{(1)}(u,\tilde{k}) +\left( \frac{2  r_+^{2}\psi_{1}\varphi A^{(1)}_{t}}{u^{2}f(u)}  +\frac{2 \tilde{a}^{(0)}_t r_+^{2}}{u^{2}f(u)} \varphi\psi_1 -2\mathcal{H}_{c2}x \tilde{a}^{(0)}_y \psi_1\right)\label{eq56}.
\end{eqnarray} 

In order to proceed further, we define the inner product between two states $ b_{i} $ and $ b_{j} $ satisfying the boundary condition (\ref{eq42}) as,
\begin{eqnarray}
\langle b_{i}| b_{j}\rangle = \int_{0}^{1}\frac{du}{u^{2}}b_{i}^{\ast}(u)b_{j}(u).\label{eq57}
\end{eqnarray} 
From the above definition (\ref{eq57}) one can easily note that the operator $ \mathcal{D}_{b} $ is Hermitian.

With the above definition (\ref{eq57}) in hand and taking the inner product of $ \psi_{1} $ with (\ref{eq56}) we note that,
\begin{eqnarray}
-\tilde{k}_{1}^{2} \langle \psi_{1}|\psi_{1} \rangle = \left\langle \psi_{1}\mid \frac{2  r_+^{2}\psi_{1}\varphi A^{(1)}_{t}}{u^{2}f(u)}\right\rangle -\frac{1}{2} \int_{0}^{1}du \left(\frac{d\tilde{a}^{(0)}_t }{du}\right)^{2}-\frac{4\kappa \mathcal{H}_{c2} }{r_{_+}}\int_{0}^{1}du \tilde{c}^{(0)}\partial_u \tilde{a}^{(0)}_t\nonumber\\
-2\mathcal{H}_{c2}x \int_{0}^{1}\frac{du}{u^{2}}\tilde{a}^{(0)}_y \psi_1^{2}\label{eq58}
\end{eqnarray}
where we have used the boundary condition (\ref{eq42}) and the fact that $ \mathcal{D}_{b}\psi_{1}=0 $. Also using the first equation in (\ref{eq56}), it is quite trivial to show that,
\begin{eqnarray}
\left\langle \psi_{1}\mid \frac{2 \tilde{a}^{(0)}_t r_+^{2}}{u^{2}f(u)} \varphi\psi_1\right\rangle &=& \frac{1}{2}\int_{0}^{1}du \frac{4r_+^{2}\varphi \psi_1^{2}}{u^{4}f(u)}\tilde{a}^{(0)}_t\nonumber\\
&=&-\frac{1}{2} \int_{0}^{1}du \left(\frac{d\tilde{a}^{(0)}_t }{du}\right)^{2}-\frac{4\kappa \mathcal{H}_{c2} }{r_{_+}}\int_{0}^{1}du \tilde{c}^{(0)}\partial_u \tilde{a}^{(0)}_t
\end{eqnarray}
where the boundary terms have been dropped out as they give rise to zero contribution when evaluated at the boundary.

Let us now consider the first term on the r.h.s. of (\ref{eq58}). Substituting (\ref{eq43}) in to the equation for the scalar field ($ \Psi $) in (\ref{eq41}) we note that,
\begin{eqnarray}
\mathcal{D}_{b} \psi_{2}=-\frac{2  r_+^{2}\psi_{1}\varphi A^{(1)}_{t}}{u^{2}f(u)}.\label{Eq107}
\end{eqnarray}
Using the above relation (\ref{Eq107}) it is now quite trivial to show that the first term on the r.h.s of (\ref{eq58}) vanishes identically,
\begin{eqnarray}
\left\langle \psi_{1}\mid \frac{2  r_+^{2}\psi_{1}\varphi A^{(1)}_{t}}{u^{2}f(u)}\right\rangle =-\left\langle \mathcal{D}_{b}\psi_{1}\mid\psi_{2}\right\rangle =0.
\end{eqnarray}

Finally Wick rotating back to the original wave vector and considering the $ \varepsilon_{T}\rightarrow 0 $ limit one finds,
\begin{eqnarray}
 -\tilde{k}_{1}^{2} = \frac{\mathcal{P}}{\mathcal{M}}\label{Eq109}
\end{eqnarray} 
where the coefficients $ \mathcal{P} $ and $ \mathcal{M} $ reads as,
\begin{eqnarray}
\mathcal{P} &=& \frac{1}{2} \int_{0}^{1}du \left(\frac{d\tilde{a}^{(0)}_t }{du}\right)^{2}+2\mathcal{H}_{c2}x \int_{0}^{1}\frac{du}{u^{2}}\tilde{a}^{(0)}_y \psi_1^{2}+\frac{4\kappa \mathcal{H}_{c2} }{r_{_+}}\int_{0}^{1}du \tilde{c}^{(0)}\partial_u \tilde{a}^{(0)}_t\nonumber\\
\mathcal{M} &=&\int_{0}^{1}\frac{du}{u^{2}} \psi_1^{2}.
\end{eqnarray}

Finally substituting (\ref{Eq109}) in to the last relation of (\ref{eq55}) we note that the expression for the coherence length turns out to be,
\begin{eqnarray}
\xi = \sqrt{\frac{\mathcal{M}}{\mathcal{P}}}\left( 1-\frac{T}{T_c}\right)^{-1/2}\propto \left( 1-\frac{T}{T_c}\right)^{-1/2}.\label{EQ111}
\end{eqnarray}

This result (\ref{EQ111}) that has been obtained in the case for the Maxwell-CS vortices in fact agrees well to that with the result of the standard GL theory for type II superconductors and indicates the onset of a second order phase transition near the critical point. Although the qualitative behaviour of the coherence length does not change, the only difference that appears here is in the numerical pre-factor where the effect of the CS coupling ($ \kappa $) is explicitly present in the coefficient ($ \mathcal{P} $) sitting at the denominator of that numerical factor.

\subsection{London equation}

In this section we compute the magnetic penetration depth ($ \delta $) as well as the GL coefficient ($ \Bbbk $) for the Maxwell-CS type vortices in ($ 2+1 $) dimensions in the presence of a homogeneous external magnetic field perpendicular to the two dimensional hyper surface. We perform our analysis close to the origin $ x=0 $ where one can effectively ignore the coupling between the gauge field fluctuations and that of the scalar field.

Let us choose the following ansatz,
\begin{eqnarray}
\delta A_{y}(u,x)=a_y (u)x.\label{Eq112}
\end{eqnarray}

Substituting the above ansatz (\ref{Eq112}) in to (\ref{Eq30}), it is quite trivial to show that $ a_y $ satisfies the following equation namely,
\begin{eqnarray}
 \left[ \partial_u \left(u^{2}f(u)\partial_u \right)-\frac{2r_+^{2}\Psi^{2}}{u^{2}}\right]a_y- \frac{4 b r_+^{2}\mathcal{H}_{c2}}{u^{2}}\Psi= 0.\label{Eq113}
\end{eqnarray}

As we have done in the previous sections, we solve the above equation (\ref{Eq113}) considering the following perturbative expansion, 
\begin{eqnarray}
a_y = a^{(0)}_{y}+\varepsilon_{T} a^{(1)}_{y}+\mathcal{O}(\varepsilon_{T}^{2}).\label{Eq114}
\end{eqnarray}

Now we need to know the corresponding $\varepsilon_{T}$ expansion for the scalar perturbation $ b (u)$. Substituting (\ref{Eq114}) in to (\ref{eq49}) with the l.h.s equal to zero\footnote{The l.h.s turns out to be zero because of the fact that we are expanding around a small neighbourhood of $ x=0 $ for which $ e^{ik.x}\approx 1 $.}, one can have the following perturbative expansion for the fluctuation $ b(u) $ namely,
\begin{eqnarray}
b = \psi_{1}+\sqrt{\varepsilon_{T}}b^{(1)}+\mathcal{O}(\varepsilon_{T})\label{EQ115}
\end{eqnarray}

Substituting (\ref{Eq114}), (\ref{EQ115}) and the $\varepsilon_{T}$ expansion of the scalar field ($ \Psi $) (see (\ref{eq43})) in to (\ref{Eq113}) we arrive at the following set of equations,
\begin{eqnarray}
\partial_u \left(u^{2}f(u)\partial_u \right)a^{(0)}_{y}&=& 0\nonumber\\
\partial_u \left(u^{2}f(u)\partial_u \right)a^{(1)}_{y}-\frac{2r_+^{2}\psi^{2}_{1}}{u^{2}}a^{(0)}_{y}- \frac{4 b^{(1)} r_+^{2}\mathcal{H}_{c2}}{u^{2}}\psi_{1}&=& 0.\label{Eq116}
\end{eqnarray}

Solutions corresponding to (\ref{Eq116}) could be formally expressed as,
\begin{eqnarray}
 a^{(0)}_{y} &=& \mathcal{C}^{(0)} = const.\nonumber\\
  a^{(1)}_{y}&=& \mathcal{C}^{(1)}-2 \mathcal{C}^{(0)} \int_{0}^{u}\frac{du''}{(1-u''^{3})}\int^{1}_{u''}\frac{du'}{u'^{2}}\psi_{1}(u')\left(\psi_{1}(u')+\frac{2b^{(1)}(u')\mathcal{H}_{c2}}{\mathcal{C}^{(0)}} \right)\label{Eq117} 
\end{eqnarray}
where $\mathcal{C}^{(0)}$ and $ \mathcal{C}^{(1)} $ are some arbitrary constants. At this stage our goal would be to solve $ b^{(1)}(u) $ perturbatively in the CS coupling ($ \kappa $). This could be done by using the following steps.

First of all note that in the neighbourhood of $ x=0 $ the perturbations could be expressed as,
\begin{eqnarray}
a_t (u,k)e^{ik.x}&\rightarrow & a_t (u)\nonumber\\
c (u,k)e^{ik.x}&\rightarrow & c (u).\label{Eq118}
\end{eqnarray} 

Using the above fact (\ref{Eq118}), from (\ref{eq46}) one can note that the zeroth order fluctuations satisfy the following equations namely\footnote{Consider the first and the last equations in (\ref{eq46}) with $ k=0 $ and $ \Psi =0 $.},
\begin{eqnarray}
 u^{2}f(u)\partial_u^{2} a^{(0)}_t +\frac{8\kappa u^{2} f(u)}{r_+}\mathcal{H}_{c2}\partial_u c^{(0)} &=& 0\nonumber\\
 \partial_u \left(f(u)\partial_u \right)c^{(0)}(u)+\frac{4\kappa}{r_+}\mathcal{H}_{c2}\partial_u a^{(0)}_t & =& 0.\label{Eq119}
\end{eqnarray}

One can go a step further in order to solve the above set of equations (\ref{Eq119}) perturbatively in $ \kappa $ namely,
\begin{eqnarray}
a^{(0)}_t = a^{(0)(\kappa^{(0)})}_t + \kappa a^{(0)(\kappa^{(1)})}_t\nonumber\\
c^{(0)} = c^{(0)(\kappa^{(0)})} + \kappa c^{(0)(\kappa^{(1)})}\label{Eq120}
\end{eqnarray}
which satisfy the following two equations,
\begin{eqnarray}
u^{2}f(u)\partial_u^{2} a^{(0)(\kappa^{(0)})}_t &=& 0 \nonumber\\
u^{2}f(u)\partial_u^{2} a^{(0)(\kappa^{(1)})}_t+\frac{8 u^{2} f(u)}{r_+}\mathcal{H}_{c2}\partial_u c^{(0)(\kappa^{(0)})} &=&0. 
\end{eqnarray}

Finally substituting (\ref{EQ115}) in to the second equation of (\ref{eq46}) one can note down the equation corresponding to $ b^{(1)} $ (near $ x=0 $) which turns out to be,
\begin{eqnarray}
\mathcal{D}_{b} b^{(1)}(u) +\frac{2 r_+^{2}a^{(0)}_t }{u^{2}f(u)} \varphi\psi_1 -2\mathcal{H}_{c2}x a^{(0)}_y \psi_1 = 0.\label{Eq122}
\end{eqnarray}
As it is quite evident from (\ref{Eq120}) that $ a^{(0)}_t $ can be solved perturbatively in the CS coupling ($ \kappa $), therefore one may also write down a solution for the above equation (\ref{Eq122}) perturbatively in $ \kappa $ namely,
\begin{eqnarray}
b^{(1)} = b^{(1)(\kappa^{(0)})} + \kappa b^{(1)(\kappa^{(1)})}+\mathcal{O}(\kappa^{2}).
\end{eqnarray}
where $ b^{(1)(\kappa^{(0)})} $ and $ b^{(1)(\kappa^{(1)})} $ satisfy the following set of equations namely,
\begin{eqnarray}
\mathcal{D}_{b} b^{(1)(\kappa^{(0)})}(u) +\frac{2 r_+^{2}\varphi\psi_1 }{u^{2}f(u)}a^{(0)(\kappa^{(0)})}_t  -2\mathcal{H}_{c2}x a^{(0)}_y \psi_1 &=& 0\nonumber\\
\mathcal{D}_{b} b^{(1)(\kappa^{(1)})}(u) +\frac{2 r_+^{2}\varphi\psi_1 }{u^{2}f(u)}a^{(0)(\kappa^{(1)})}_t  -2\mathcal{H}_{c2}x a^{(0)}_y \psi_1 &=& 0\label{eq77}.
\end{eqnarray}

Adding the two pieces in (\ref{Eq117}) from (\ref{Eq114}) we have,
\begin{eqnarray}
a_y (u)=\mathcal{C}^{(0)} + \varepsilon_{T} \left[ \mathcal{C}^{(1)}-2 \mathcal{C}^{(0)} \int_{0}^{u}\frac{du''}{(1-u''^{3})}\int^{1}_{u''}\frac{du'}{u'^{2}}\psi_{1}(u')\left(\psi_{1}(u')+\frac{2b^{(1)}(u')\mathcal{H}_{c2}}{\mathcal{C}^{(0)}} \right)\right] + \mathcal{O}(\varepsilon_{T}^{2}).\nonumber\\
\end{eqnarray}

In order to proceed further, as a next step we choose $ \mathcal{C}^{(0)}= \mathcal{H}_{c2} $ and $ \mathcal{C}^{(1)}=0 $. With this choice from (\ref{Eq112}) we get,
\begin{eqnarray}
\delta A_y (u,x)&=&\delta A^{(0)}_y (x)\left( 1 -2\varepsilon_{T}  \int_{0}^{u}\frac{du''}{(1-u''^{3})}\int^{1}_{u''}\frac{du'}{u'^{2}}\psi_{1}(u')\left(\psi_{1}(u')+2b^{(1)}(u') \right) \right)+ \mathcal{O}(\varepsilon_{T}^{2})\nonumber\\
 &\approx&\delta A^{(0)}_y (x)\left( 1 -2\varepsilon_{T}  u\int^{1}_{u''}\frac{du'}{u'^{2}}\psi_{1}(u')\left(\psi_{1}(u')+2b^{(1)}(u') \right) \right)+ \mathcal{O}(\varepsilon_{T}^{2})\label{eq79}
\end{eqnarray}
where $ \delta A^{(0)}_y (x)=  \lim_{u\rightarrow 0} \delta A_y (u,x)=\mathcal{H}_{c2}  x $.

Finally, using (\ref{eq5}), (\ref{eqJ}) and (\ref{eq79}) and considering our system close to the critical point ($ T\sim T_c $) we obtain,
\begin{eqnarray}
\langle J_{y}(x)\rangle_{u\rightarrow 0}=- \frac{8 \pi \varepsilon_{T} T_c}{3}\delta A^{(0)}_y (x)\int^{1}_{0}\frac{du}{u^{2}}\mathcal{G}(u,\kappa)+ \mathcal{O}(\varepsilon_{T}^{2})\label{eq80}
\end{eqnarray}
where,
\begin{eqnarray}
\mathcal{G}(u,\kappa)= \psi^{2}_{1}(u)+2\psi_{1}(u)b^{(1)(\kappa^{(0)})}(u)+2\kappa \psi_{1}(u)b^{(1)(\kappa^{(1)})}(u).
\end{eqnarray}

Before we proceed further, it is now time to explore the behaviour of the above integral in (\ref{eq80}) near the boundary of the $ AdS_{4} $. First of all note that near the boundary of the $ AdS_{4} $ (for $ m^{2}=-2 $) the scalar field has the following behaviour namely\footnote{Since we are close to the critical temperature, therefore $  \langle\mathcal{O}\rangle $ plays the role of the order parameter for the boundary theory.},
\begin{eqnarray}
\psi_{1}\sim \langle\mathcal{O}\rangle u^{2}.
\end{eqnarray}
On top of it $ \psi_{1} $ is perfectly regular near the horizon $ u=1 $. All these facts suggest that the integral above in (\ref{eq80}) is perfectly well defined and yields a finite number.


Using all these facts, from (\ref{eq80}) near the boundary of the $ AdS_{4} $ we have,
\begin{eqnarray}
\langle J_{y}(x)\rangle_{u\rightarrow 0}=- \frac{8 \pi \varepsilon_{T} T_c }{3}\langle\mathcal{O}\rangle^{2}\delta A^{(0)}_y (x)F(\Delta , \kappa)+ \mathcal{O}(\varepsilon_{T}^{2})\label{eq84}
\end{eqnarray}
where $ F(\Delta , \kappa)=\int^{1}_{0}\frac{du}{u^{2}}\mathcal{G}(u,\kappa)=\mathcal {F}_{1}(\Delta)+\kappa\mathcal {F}_{2}(\Delta) $ where $ {F}_{1}(\Delta) $ and $ \mathcal {F}_{2}(\Delta) $ are purely some $ c $ numbers.

It is quite interesting to note that Eq.(\ref{eq84}) is remarkably identical to that of the so called London equation namely,
\begin{eqnarray}
\textbf{J}=-\frac{e_{\ast}^{2}}{m_{\ast}}\psi^{2}\textbf{A}=-e_{\ast}n_{s}\textbf{A}\label{eq85}.
\end{eqnarray}

One crucial point that has to be noted at this stage is the following: The gauge field $ \textbf{A} $ that appears in the London equation above in (\ref{eq85}) has two parts in it, one is the spatial average of the \textit{microscopic} gauge field and the other contribution comes from the external source which may be regarded as some macroscopic gauge field. Since in the holographic picture the gauge fields do not have any dynamics at the boundary, therefore the gauge field ($ \delta A^{(0)}_y (x) $) that is appearing in (\ref{eq84}) should be strictly considered to be the macroscopic gauge field for the Maxwell-CS vortices. The current that is generated due to the external magnetic field at the boundary of the $ AdS_{4} $ eventually cancels out the effect of the applied magnetic field itself. As a consequence of this one can in fact reproduce London equation using the holographic techniques. 

Comparing (\ref{eq84}) and (\ref{eq85}) the number density of the super fluid particles turns out to be,
\begin{equation}
n_{s}= \frac{8 \pi \varepsilon_{T} T_c }{3}\langle\mathcal{O}\rangle^{2}F(\Delta , \kappa).\label{Eq132}
\end{equation}
From (\ref{Eq132}) the magnetic penetration depth ($ \delta $) turns out to be,
\begin{eqnarray}
\delta = \frac{1}{\sqrt{n_s}}=\left(\frac{3}{8 \pi \varepsilon_{T}T_c F(\Delta , \kappa)} \right)^{1/2}\langle\mathcal{O}\rangle^{-1}.\label{Eq133}
\end{eqnarray}

Finally using (\ref{EQ111}) and (\ref{Eq133}), the GL coefficient ($ \Bbbk $) for the Maxwell-CS type vortices turns out to be,
\begin{eqnarray}
\Bbbk =\frac{\delta}{\xi}= \left(\frac{3 \mathcal{P}}{8 \mathcal{M}\pi T_c F(\Delta , \kappa)} \right)^{1/2}\langle\mathcal{O}\rangle^{-1}.
\end{eqnarray}

\section{Summary and final remarks}

In the present paper, based on the $ AdS_{4}/CFT_{3} $ duality we have systematically addressed the issue whether there could be a dual gravitational description for the abelian Chern-Simons (CS) Higgs model that has been widely studied in ($ 2+1 $) dimensions. In order to answer this question we have tried to construct our theory based on an approach that is quite similar in spirit to that of the standard Ginzburg-Landau (GL) approach for type II superconductors. 
As a first step of our analysis, we have computed the electromagnetic current associated with the Maxwell-CS Higgs vortices in ($ 2+1 $) dimensions. It is observed that the current could be expressed as a local function of the vortex solution in the closed neighbourhood of the vortex \cite{Maeda:2009vf},\cite{ref24}. It is also found that apart from having the usual GL piece the local form of the current also contains an additional correction term depending on the CS coupling ($ \kappa $). 

In the next step of our analysis, we have reproduced this current based on the standard holographic prescription. In order to compute the current holographically we have considered an abelian Higgs model coupled to gravity in $ AdS_{4} $ space time in the presence of a $ \theta F\wedge F $ term \cite{ref36}. It has been observed that using this gravitational model one can in fact reproduce an identical local expression for the current that has been obtained for the Maxwell-CS type vortices in ($ 2+1 $) dimensions. This remarkable agreement motivates us to claim that the gravitational theory mentioned above could be regarded as the dual description for the Maxwell-CS type vortices that naturally emerges from the abelian CS Higgs model in ($ 2+1 $) dimensions. Finally, we have explored this duality a bit further in order to compute the coherence length ($ \xi $) as well as the GL coefficient ($ \Bbbk $) associated with the Maxwell-CS type vortices in ($ 2+1 $) dimension. It is observed that all these entities exhibit almost identical qualitative features as that is found for ordinary type II superconductors which suggests that the Maxwell-CS vortices are qualitatively quite similar to that of the ordinary type II vortices.   

Before we conclude this article, we would like to make some general comments regarding the further applicability of the method that has been developed throughout this paper. One immediate application could be the construction of a dual gravitational description for the non abelian CS vortices. Hopefully using the machinery developed in this article one would be able to construct a correct gravitational description that would eventually give rise some meaningful expression for the current as well as the other relevant entities for the non abelian CS vortices.

{\bf {Acknowledgements :}}
Author would like to thank IISER (Bhopal) for the financial support at the initial stage of this work. The author also would like to acknowledge the financial support from CHEP, Indian Institute of Science, Bangalore.\\
\\\ \\\
 
\appendix

\noindent
{\bf \large Appendix}
\section*{Calculation of $ j_{i}^{(0)} $} 
From (\ref{Eq21}) we note that,
\begin{eqnarray}
\Psi_{1}^{(0)}=e^{-\frac{x^{2}}{2\xi^{2}}}\vartheta_{3}(v,\tau).
\end{eqnarray}

Consider $ i=x $ for which the current turns out to be,
\begin{eqnarray}
j_{x}^{(0)}&=& i\left( \Psi_{1}^{(0)} \nabla_{x} \Psi_{1}^{(0)\dagger}-\Psi_{1}^{(0)\dagger} \nabla_{x} \Psi_{1}^{(0)}\right) - \kappa \mid \Psi_{1}^{(0)}(\textbf{x})\mid^{2}A^{(0)(\kappa^{(1)})}_{x}+\mathcal{O}(\kappa^{2}).\label{Eq136}
\end{eqnarray}

Consider the first term in the parenthesis namely,
\begin{eqnarray}
\left( \Psi_{1}^{(0)} \nabla_{x} \Psi_{1}^{(0)\dagger}-\Psi_{1}^{(0)\dagger} \nabla_{x} \Psi_{1}^{(0)}\right)=\frac{i e^{-\frac{x^{2}}{\xi^{2}}}}{a_y}\left[\frac{\partial \vartheta^{\dagger}_{3}}{\partial v}\vartheta_{3}+ \vartheta_{3}^{\dagger}\frac{\partial \vartheta_{3}}{\partial v}\right] = \frac{i e^{-\frac{x^{2}}{\xi^{2}}}}{a_y}\frac{\partial}{\partial v}(\vartheta^{\dagger}_{3}\vartheta_{3}).\label{Eq137}
\end{eqnarray}

On the other hand,
\begin{eqnarray}
\frac{\partial}{\partial y}\mid \Psi_{1}^{(0)}(\textbf{x})\mid^{2}= e^{-\frac{x^{2}}{\xi^{2}}}\frac{\partial}{\partial y}(\vartheta^{\dagger}_{3}\vartheta_{3})=\frac{e^{-\frac{x^{2}}{\xi^{2}}}}{a_y}\frac{\partial}{\partial v}(\vartheta^{\dagger}_{3}\vartheta_{3}).\label{Eq138}
\end{eqnarray}

Substituting (\ref{Eq13}), (\ref{Eq137}) and (\ref{Eq138}) in to (\ref{Eq136}) we finally obtain,
\begin{eqnarray}
j_{x}^{(0)} = - \epsilon_{x}^{y}\partial_{y}\mid \Psi_{1}^{(0)}(\textbf{x})\mid^{2}+\kappa \Pi(\Delta)\epsilon_{x}^{y}\partial_{y}\mid \Psi_{1}^{(0)}(\textbf{x})\mid^{2}+\mathcal{O}(\kappa^{2}).
\end{eqnarray}

Following the same steps as mentioned above one can also compute the $ y $ component of the current namely $ j_{y}^{(0)} $.


\end{document}